\documentclass[10pt]{iopart}
\usepackage{graphicx,amssymb,amsfonts,units,bm,amsthm,yhmath,hyperref}

\begin{document}

\title{The non-commutative $n${\scriptsize th}-Chern number ($n \geq 1$)}

\author{Emil Prodan}

\address{Department of Physics, Yeshiva University, New York, NY 10016, USA \\
\href{mailto:prodan@yu.edu}{prodan@yu.edu}}

\author{Bryan Leung}

\address{Department of Physics \& Astronomy, Rutgers University, Piscataway, NJ 08854, USA \\
\href{mailto:bryleung@physics.rutgers.edu}{bryleung@physics.rutgers.edu}}

\author{Jean Bellissard}

\address{Georgia Institute of Technology, School of Mathematics, Atlanta GA 30332-0160, USA \\
\href{mailto:jeanbel@math.gatech.edu}{jeanbel@math.gatech.edu}}

\begin{abstract}
The theory of the higher Chern numbers in the presence of strong disorder is developed. Sharp quantization and homotopy invariance conditions are provided. The relevance of the result to the field of strongly disordered topological insulators is discussed.
\end{abstract}

\maketitle

\section{Introduction}

The non-commutative 1{\scriptsize st}-Chern number \cite{BELLISSARD:1994xj} can be viewed as a generalization of the TKNN invariant \cite{ThoulessPRL1982vh} to the case of aperiodic systems. While the TKNN invariant is protected by a spectral gap and can be defined only for uniform magnetic fields satisfying the ``rational-flux" condition, the non-commutative 1{\scriptsize st}-Chern number is protected by a mobility gap and can be defined for arbitrary uniform magnetic fields. These special properties of the 1{\scriptsize st}-Chern number played a decisive role in our understanding of the Integer Quantum Hall Effect (IQHE) \cite{BELLISSARD:1994xj}. In the new field of topological insulators, the non-commutative 1{\scriptsize st}-Chern number was successfully used to characterize and compute the phase diagram of strongly disordered 2-dimensional Chern insulators \cite{Prodan2010ew,ProdanJPhysA2011xk}. The non-commutative 1{\scriptsize st}-Chern number also played an instrumental role in the definition of the non-commutative spin-Chern number \cite{Prodan:2009oh}. The latter proved to be an extremely effective tool in the characterization and computation of the phase diagram of strongly disordered 2-dimensional quantum spin-Hall insulators \cite{ProdanJPhysA2011xk,Prodan2011vy,XuPRB2012vu}.

Recent developments in the field of topological insulators brought to light a new class of 3-dimensional materials called strong topological insulators \cite{Moore:2007ew,Fu:2007vs,Hsieh:2008vm}. Many such materials have been already fabricated and finely characterized in laboratories \cite{ZHassanRevModPhys2010du,QiRMP2011tu}. Their hallmark characteristics are a topological magneto-electric response in the bulk \cite{QiPRB2008ng} and formation of metallic states at the surface \cite{Moore:2007ew,Fu:2007vs}. It is believed that these characteristics remain stable as long as the time-reversal invariance is strictly enforced and the insulating character is maintained in the bulk of the material. The theory of strong topological insulators is on a rigorous footing for perfectly periodic crystals \cite{Fu:2007ti,Moore:2007ew,Roy:2009am}, but little progress has been made for strongly disordered crystals. Some of the important theoretical open questions for the latter case are: Do the topological invariants, defined for the periodic case, continue to make sense at strong disorder when the Fermi level is embedded in dense localized spectrum? Do bulk extended states exist at strong disorder, like in the IQHE? Are the metallic surface states robust against strong disorder? Some of these open questions have been numerically investigated \cite{LeungPRB2012vb} and the available results hint to positive answers. On the analytical front, it was recently noted \cite{LeungJPA2012er} that a rigorous theory of the non-commutative 2{\scriptsize nd}-Chern number in dimension four could provide a rigorous answer to the second open question. This is the main motivation for the present work. 

The 2{\scriptsize nd}-Chern number enters the stage in the following way. Let us consider first the perfectly periodic 3-dimensional insulators. So let there be such a quantum system, described by a Hamiltonian $H_{\bm \lambda}$ which depends on a set of parameters ${\bm \lambda}=\{\lambda_1,\ldots\}$. We assume a system of units in which $h=e=1$. The key physical property to consider is the isotropic part of the magneto-electric response of the system:
\begin{equation}
\alpha=\frac{1}{3}\sum_{i=1}^3 \frac{\partial P_i}{\partial B_i} = \frac{1}{3} \sum_{i=1}^3 \frac{\partial M_i}{\partial E_i},
\end{equation}
where ${\bm P}$ and ${\bm M}$ are the vectors of electric polarization and magnetization, respectively, and ${\bm B}$ and ${\bm E}$ are the magnetic and electric fields, respectively. Let ${\bm \lambda}_1$ and ${\bm \lambda}_2$ be two points in the parameter space defining two Hamiltonians that are time-reversal symmetric, and consider a path $\gamma$, not necessarily time-reversal invariant, connecting ${\bm \lambda}_1$ and ${\bm \lambda}_2$ such that the Fermi level of $H_{\bm \lambda}$ is in a spectral gap for all  ${\bm \lambda}\in \gamma$. In this conditions, the variation of the isotropic magneto-electric response along $\gamma$ is given by \cite{QiPRB2008ng, EssinPRL2009bv,Hughes2010gh,EssinPRB2010ls,MalashevichNJP2010bv}:
\begin{equation}\label{ME1}
\Delta \alpha (\gamma) = \frac{1}{2}C_2(\gamma-\theta \gamma).
\end{equation}
Here, $C_2(\gamma-\theta \gamma)$ is the 2{\scriptsize nd}-Chern number of the vector bundle of the occupied electron states (see Eq.~\ref{Chern1}), defined over the manifold generated by the closed path $\gamma-\theta \gamma$ times the 3-dimensional Brillouin torus. The symbol $\theta$ represents the time-reversal operation in the parameter space of the Hamiltonian. Choosing a standard reference system, it follows from Eq.~\ref{ME1} and other considerations that the time-reversal insulators fall into two topologically distinct classes, according to the integer or half-integer character of $\alpha$ (a property that is path-independent). This is the well established ${\bm Z}_2$ classification of periodic, time-reversal symmetric insulators \cite{Moore:2007ew,Fu:2007ti,QiPRB2008ng,Roy:2009am}. Two systems from the two different classes cannot be connected by a time-reversal path (i.e. $\theta \gamma = \gamma$) without continuum energy spectrum crossing the Fermi level. 

Starting from the non-commutative formula for electric polarization reported in Ref.~\cite{Schulz-BaldesCMP2013gh}, the magneto-electric response was computed in the presence of disorder and under the gap condition, and the result was \cite{LeungJPA2012er}:
\begin{equation}
\Delta \alpha (\gamma) = \frac{1}{2}C_2(\gamma-\theta \gamma),
\end{equation}
where $C_2(\gamma-\theta \gamma)$ is the non-commutative 2{\scriptsize nd}-Chern number, as defined in Eq.~\ref{Chern4}, over the non-commutative manifold generated by the closed loop $\gamma-\theta \gamma$ times the non-commutative Brillouin torus of the 3-dimensional aperiodic crystal. Our hope is that the theory of the higher non-commutative Chern numbers developed by the present work will enable new progress on the classification of the strong topological insulators in the presence of disorder, that goes beyond the limitations imposed by the spectral gap condition. It is important to note that the non-commutative formulas established in this work have also a practical value, as they can be efficiently and accurately evaluated on a computer, using for example the methods developed in Ref.~\cite{ProdanAMRX2013bn}.

We now discuss the structure of the paper and our main result. Given the potentially interesting physical applications, we decided to put our result in a context familiar to the condensed matter theorists, namely, to work with the projectors onto the occupied electron states of typical disordered $2n$-dimensional lattice-models. These models are progressively discussed  in Section 2, starting from the simplest case of translational invariant ones and ending with the general class of homogeneous lattice systems, which includes the disordered quantum lattice-models under uniform magnetic fields. In Eq.~\ref{Chern1} we present the familiar expression of the $n${\scriptsize th}-Chern number over the $2n$-dimensional Brillouin torus of a perfectly periodic insulator. Its equivalent real-space representation is given in Eq.~\ref{Chern2} and, as we shall see, this formula leads to a canonical extrapolation of the $n${\scriptsize th}-Chern number formula from the periodic to the aperiodic case. The formula for the latter case is given in Eq.~\ref{Chern3}. It is precisely the expression written in Eq.~\ref{Chern3} that was found to be connected to the magneto-electric response of 3-dimensional topological insulators. As such, we take Eq.~\ref{Chern3} as the definition of the $n${\scriptsize th}-Chern number for aperiodic lattice systems.

The bulk of the paper is contained in Sections 4 and 5, and is dedicated to understanding the topological properties of the $n${\scriptsize th}-Chern number, namely, its quantization and homotopy invariance, together with the optimal conditions when these happen. The natural framework for the analysis is the non-commutative Brillouin torus $({\cal A},{\cal T},\partial)$ of the aperiodic homogeneous lattice system. An equivalent representation of the $n${\scriptsize th}-Chern number, this time written over this non-commutative manifold, is reported in Eq.~\ref{Chern4} (see also below). We call this formula the non-commutative $n${\scriptsize th}-Chern number. Our main results can be summarized as follows.\medskip

\noindent{\it Summary of the main results.} A) Consider the settings and the notations introduced in Sections 2 and 3. In particular, let $({\cal A},{\cal T},\partial)$ be the non-commutative Brillouin torus of a disordered homogeneous lattice system,  and let $h \in {\cal A}$ be the element generating the covariant family of disordered lattice Hamiltonians, and $p=\chi_{(-\infty,\epsilon_F]}(h)$ be the projector onto the occupied states. Based on the magneto-electric response of strong topological insulators, we propose the following definition of the non-commutative $n${\scriptsize th}-Chern number:
\begin{equation}\label{Chern0}
C_n \stackrel{\text{\tiny def}}{=}  \frac{(2 \pi \imath)^n}{n!} \sum_{\sigma}(-1)^\sigma {\mathcal T} \left ( p \prod_{i=1}^{2n} \partial_{\sigma_i}p\right ).
\end{equation}

B) Let $\gamma_1, \ldots, \gamma_{2n}$ be an irreducible representation of the Clifford algebra $Cl_{2n,0}$ in the finite $2^n$-dimensional Hilbert space $\mathrm{Cliff}(2n)$. Let ${\mathcal H}=\ell^2(\mathbb{Z}^{2n},\mathbb{C}^Q)\otimes \mathrm{Cliff}(2n)$ be the augmentation of the physical Hilbert space by this space, and let $D=\sum_{i=1}^{2n}X^i \gamma_i$ be the Dirac operator on ${\cal H}$ and $\hat{D}=D/|D|$. Let $\omega$ be a disorder configuration and $\pi_\omega$ be the standard representation of ${\cal A}$ on ${\cal H}$, and let $\pi_\omega^\pm(p)$ be the decomposition of $\pi_\omega(p)$ according to the grading induced by $\gamma_0$ on $\mathcal H$. If the localization length
\begin{equation}\label{Cond}
\Lambda_n \stackrel{\text{\tiny def}}{=} \sum_{i=1}^{2n} {\mathcal T}\left( |\partial_i p|^{2n}\right)^{\frac{1}{2n}}
\end{equation}
is finite, then, with probability one in $\omega$, the operator $\pi_\omega^-(p)\hat{D} \pi_\omega^+(p)$ is in the Fredholm class and, with probability one, its Fredholm index is independent of $\omega$ and is equal to the non-commutative $n${\scriptsize th}-Chern number:
\begin{equation}
\mathrm{Index}\left(\pi_\omega^-(p)\hat{D} \pi_\omega^+(p)\right)=C_n.
\end{equation}
As such, the non-commutative $n${\scriptsize th}-Chern number remains constant and quantized under continuous homotopies of $p$, where the continuity is considered with respect to an appropriate Sobolev norm (see Eq.~\ref{SobolevNorm}).

C) Consider a disordered lattice system with finite hopping range. Consider continuous variations of the hopping amplitudes and assume that the localization condition of Eq.~\ref{FracMom1} at the Fermi level holds (this is the Aizenman-Molchanov bound on the fractional powers of the resolvent \cite{Aizenmann1993uf}). Then the projector $p$ is in the Sobolev space mentioned at point B), and $p$ varies continuously with respect to the Sobolev norm. As a consequence, $C_n$ stays constant and quantized.\medskip

For part B), our proof follows closely the main steps from the work on the non-commutative 1{\scriptsize st}-Chern number \cite{BELLISSARD:1994xj}. The key technical points are a generalization of a pivotal identity (see Eq.~\ref{Identity}) due to Connes in 2-dimensions \cite{CONNES:1985cc}, and the Dixmier trace calculation in higher dimensions. For part C), we mainly follow Ref.~\cite{Richter2001jg} where the arguments were developed for the 1{\scriptsize st}-Chern number. Addressing to the people more familiar with the subject, we want to mention that the key conceptual element that led to our result was the construction of the correct Fredholm module. For the non-commutative 1{\scriptsize st}-Chern number \cite{BELLISSARD:1994xj}, the unitary transformation $\psi \rightarrow \frac{x^1+\imath x^2}{|{\bm x}|}\psi$, describing the effect of a Dirac flux-tube threaded through the lattice, played an essential role in the construction of the Fredholm module. Under this physical picture, however, it is difficult to see how to extend the analysis to higher dimensions. What is the unitary transformation that we need to consider in higher dimensions and what is being threaded through the lattice? One useful observation was that the same Fredholm module emerges if one takes $F$ to be the multiplication with $(\sigma_1 x^1+\sigma_2 x^2)/|{\bm x}|$, where $\sigma$'s are the usual Pauli matrices (here we follow the notation from Connes' book \cite{Connes:1994wk}). This is the same to say that $F=D/|D|$, where $D$ is the Dirac operator on the Hilbert space of the 2-dimensional model augmented by a representation of the appropriate Clifford algebra. We recall that here we are investigating the geometry of the Brillouin torus, but we work in the real-space representation and that is why the Dirac operator takes this form (with multiplications instead of derivations). Now, the Dirac operator, or better said Dirac-like operators, play an essential role in the local non-commutative index formula \cite{ConnesGFA1995re}, so it was natural to enquire if the extension to higher dimensions is simply given by a Fredholm module defined by $F=D/|D|$, with $D={\bm X}\cdot{\bm \gamma}$, where $\gamma_i$'s generate an irreducible representation of the appropriate Clifford algebra for dimension 2n. The answer is affirmative. In fact, a spectral triple based on $D$ was already introduced in Ref.~\cite{BellissardLN2003bv}, and its relation to the work of Alain Connes was briefly described there. It will  be interesting to continue that discussion and see if one can establish a direct connection between our result and the local non-commutative index formula by Connes and Moscovici \cite{ConnesGFA1995re}. This will be deferred to future investigations.

Lastly, we would like to clarify for the reader how the Chern numbers defined for lattice-models over $\mathbb{Z}^4$ are relevant for the magneto-electric response of the 3-dimensional topological insulators. For this let us consider an interpolation between a trivial ($H_1$) and a topological ($H_2$) phase of a 3-dimensional insulator with time-reversal symmetry. Such interpolation can be constructed in a canonical way \cite{EssinPRL2009bv}: 
$$H(t)=\nicefrac{1}{2}(H_1+H_2)+\nicefrac{1}{2}\cos(t)(H_1 - H_2) + \sin(t)V, \ \ t\in [0,\pi],$$ 
where $V$ is a potential which changes sign upon a conjugation with the time-reversal operation. Like in the previous discussion, let us assume that the Fermi level $\epsilon_F$ is located in the middle of a spectral gap for all $t$'s. It is important to pin $\epsilon_F$ to zero (or any other value) using a $t$-dependent trivial shift of $H(t)$. The closed loop $\gamma - \theta \gamma$ is obtained by simply letting $t$ go from $0$ to $2\pi$. Now, this interpolating Hamiltonian acts on $\ell^2(\mathbb{Z}^3,\mathbb{C}^Q)$, but we can view it as acting on $\ell^2(\mathbb{Z}^3,\mathbb{C}^Q)\otimes L^2(S_1)$ ($S_1$ = the unit circle), by changing $t$ from a mere parameter to a genuine coordinate. Indeed, this new Hamiltonian will display an insulating gap at $\epsilon_F$ and the 2{\scriptsize nd}-Chern number for the corresponding projector is exactly the Chern number over the non-commutative torus times $S^1$ which connects to $\Delta \alpha$. But we can equivalently write $H(t)$ in the frequency domain, i.e. on $\ell^2(\mathbb{Z}^4,\mathbb{C}^Q)$ where it takes the form:
$$H= \nicefrac{1}{2}(H_1+H_2)+\nicefrac{1}{4}(T_4+T_4^{-1})(H_1 - H_2) + \nicefrac{\imath}{2} (T_4 - T_4^{-1})V,$$
with $T_4$ being the translation by one unit in the 4-th coordinate. This representation leads us directly to the 2{\scriptsize nd} non-commutative Chern defined in Eq.~\ref{Chern0} with $n=2$. The disorder in $H$ is highly anisotropic but an important point is that our theory works for such case too. 

We also want to mention that, recently, it has been demonstrated theoretically and experimentally that artificial crystals can be generated in arbitrary dimensions using cyclic pumping. For example a 2-dimensional 1{\scriptsize st}-Chern insulator was created in the lab by cyclicly pumping a 1-dimensional photonics quasicrystal \cite{KrausPRL2012hh}. In the same work, it was argued that a 2{\scriptsize nd}-Chern insulator in 4-dimensions can be generated by cyclicly pumping a 3-dimensional quasicrystal. The strategy works for regular lattices too \cite{Gomez-LeonPRL2013dd}, and now the class of the so called Floquet Topological Insulators is firmly established. When analyzed in the frequency domain, these systems lead to genuine lattice-models in dimension $d+1$, where $d$ is the spacial dimensionality of the system. As such, the theory of the higher Chern numbers developed in this work can find multiple applications in the field of topological materials.

\section{Lattice-models and their Chern numbers}\label{Heuristics}

\subsection{The classical $n${\scriptsize th}-Chern number for periodic crystals}

Let us consider a generic quantum lattice-model over $\mathbb{Z}^{2n}$, whose Hilbert space is $\ell^2(\mathbb{Z}^{2n},\mathbb{C}^Q)$, with $Q$ a strictly positive integer. Let $|{\bm x},\alpha \rangle$ be the natural basis of this space, where ${\bm x}\in \mathbb{Z}^{2n}$ is a node of the lattice and $\alpha=1,\ldots,Q$ labels the atomic or molecular orbitals associated with that node. In the absence of magnetic fields and disorder, a typical lattice Hamiltonian takes the form:  
\begin{equation}\label{LatticeHam0}
H_0=\sum_{{\bm x},\alpha}\sum_{{\bm y},\beta} t_{{\bm x}-{\bm y}}^{\alpha \beta} |{\bm x},\alpha \rangle  \langle {\bm y},\beta |.
\end{equation}
Throughout our work, we will assume that $t_{{\bm x}-{\bm  y}} \neq 0$ only if $|{\bm x}-{\bm y}|<R$ (finite hopping range), with $R$ arbitrarily large but, nevertheless, finite and fixed. It is convenient to fix the Fermi level at zero. The number of occupied electron states per volume can be controlled by introducing an additive constant to the Hamiltonian. We assume that this constant is already incorporated in Eq.~\ref{LatticeHam0}. Due to the translation invariance, the standard Bloch-Floquet transformation can be used to transfer the model over the Brillouin torus $\mathbb{T}=\mathcal{S}_1\times \ldots \times \mathcal{S}_1$, where the Hamiltonian is represented by a family of analytic $Q\times Q$ matrices $\hat{H}_{\bm k}$, ${\bm k} \in \mathbb{T}$. Here, $\mathcal{S}_1$ represents the unit circle in $\mathbb{R}^2$. 

If $H_0$ has a spectral gap at the Fermi level, then the spectra of all $\hat{H}_{\bm k}$'s are void at and near the origin, and consequently one can define the ${\bm k}$-dependent projector $\hat{P}_{\bm k}=\chi_{(-\infty,0]}(\hat{H}_{\bm k})$, which is a $Q \times Q$ matrix with entries analytic of ${\bm k}\in \mathbb{T}$. The analytic family of projectors $\{\hat{P}_{\bm k}\}_{{\bm k}\in \mathbb{T}}$ defines a vector bundle over the Brillouin torus, for which one can define the standard curvature 2-form:
\begin{equation}
{\cal F} = \hat{P}_{\bm k} \ d \hat{P}_{\bm k}\wedge d \hat{P}_{\bm k}.
\end{equation}
The classical $n${\scriptsize th}-Chern number over the Brillouin torus is given by the formula \cite{AvronCMP1989fj}:
\begin{equation}\label{Chern1}
C_{n}=\nicefrac{(-1)^n}{(2 \pi \imath)^n n!} \int_{\mathbb{T}} \ \mathrm{tr}\{{\cal F}^n\}.
\end{equation}
In Eq.~\ref{Chern1} and throughout the paper, $\mathrm{tr}\{  \}$ represents the trace over the orbital index $\alpha$ and $\imath=\sqrt{-1}$. It is a standard result in differential topology that, as long as $\hat P_{\bm k}$ remains globally smooth of ${\bm k}$, its $n${\scriptsize th}-Chern number takes integer values that remains constant under smooth deformations of the $\{\hat P_{\bm k}\}_{{\bm k}\in \mathbb{T}}$ family. In the present context, the smoothness of $\hat P_{\bm k}$ is protected by the spectral gap of $H_0$.

Using the Bloch-Floquet transformation in reverse, Eq.~\ref{Chern1} can be written in the real-space representation:
\begin{equation}\label{Chern2}
C_{n}=\nicefrac{(2 \pi \imath)^n}{n!} \sum_{\sigma}(-1)^\sigma \mathrm{Tr}\big \{ P_0 \prod_{i=1}^{2n} \big (\imath[X_{\sigma_i},P_0]\big ) \chi_{\bm 0} \big \},
\end{equation}
where $P_0$ is the spectral projector onto the occupied electron states for the total Hamiltonian: $P_0=\chi_{(-\infty,0]}(H_0)$, and $(-1)^\sigma$ is the signature of the permutation $\sigma$. Also, $X_i$, $i=1,\ldots,2n$, are the components of the position operator ${\bm X}$, and $\chi_{\bm x}=\sum_{\alpha=1}^Q|{\bm x},\alpha \rangle \langle {\bm x},\alpha|$ is the projector onto the states at site ${\bm x}$.

\subsection{A canonical extension to aperiodic crystals}

Let us first explain the class of aperiodic lattice-models we consider in this work. They include disorder and a uniform magnetic field ${\bm B}$ (parametrized as a $2n\times 2n$ antisymmetric  tensor $\hat{{\bm B}}$), in which case the lattice-Hamiltonian takes the form:
\begin{equation}\label{LatticeHamOmega}
H_\omega=\sum_{{\bm x},\alpha}\sum_{{\bm y},\beta} e^{\imath\pi({\bm x}, \hat{{\bm B}} {\bm y})}t_{{\bm x},{\bm y}}^{\alpha \beta}(\omega) |{\bm x},\alpha \rangle  \langle {\bm y},\beta |.
\end{equation}
The effect of the magnetic field is capture by the Peierls phase factor $e^{\imath\pi({\bm x}, \hat{{\bm B}} {\bm y})}$ \cite{PeierlsZPhys1933fj}, where $(,)$ is the Euclidean scalar product on $\mathbb{R}^{2n}$. We choose a system of units such that $h=e=1$. A disordered displacement of the atomic positions in a crystal introduces a random component in the hopping amplitudes, primarily because of the induced variations in the overlap integrals of the orbitals. To be explicit, we consider the following particular form:
\begin{equation}
t_{{\bm x},{\bm y}}^{\alpha \beta}(\omega)=t_{{\bm x}-{\bm y}}^{\alpha \beta}+\lambda \omega^{\alpha \beta}_{{\bm x},{\bm y}},
\end{equation}
where $\omega^{\alpha \beta}_{{\bm x},{\bm y}}$ are independent random variables, uniformly distributed in the interval $[-\frac{1}{2},\frac{1}{2}]$. The collection of all random variables $\omega=\{ \omega^{\alpha \beta}_{{\bm x},{\bm y}}\}$ can be viewed as a point in an infinite dimensional configuration space $\Omega$, which is metrizable and can be equipped with the probability measure:
\begin{equation}
d\mu(\omega)=\prod_{\alpha \beta}\prod_{{\bm x},{\bm y}} d\omega^{\alpha \beta}_{{\bm x},{\bm y}}.
\end{equation}
From now on, $\Omega$ will be viewed as a probability space. Furthermore, the natural action of the discrete $\mathbb{Z}^{2n}$ additive group on $\Omega$:
\begin{equation}
(\mathfrak{t}_{\bm a} \omega)^{\alpha \beta}_{{\bm x},{\bm y}}=\omega^{\alpha \beta}_{{\bm x}-{\bm a},{\bm y}-{\bm a}}, \ a\in \mathbb{Z}^{2n},
\end{equation}
acts ergodically and leaves $d\mu(\omega)$ invariant. If $U_{\bm a}$ denotes the magnetic translation by ${\bm a}$:
\begin{equation}
U_{\bm a}|{\bm x},\alpha \rangle = e^{ -\imath \pi ({\bm a}, \hat{{\bm B}} {\bm x})}|{\bm x}+{\bm a},\alpha \rangle,
\end{equation}
 then $\{H_\omega\}_{\omega \in \Omega}$ forms a covariant family of operators in the sense that:
\begin{equation}
U_{\bm a}H_\omega U_{\bm a}^{-1}=H_{\mathfrak{t}_{\bm a}\omega}.
\end{equation}
In these conditions, the triplet $(\Omega,\mathfrak{t},\{H_\omega\}_{\omega \in \Omega})$ defines a homogenous system, as defined in Ref.~\cite{BellissardLNP1986jf}. The results stated in the next Sections can be adapted to work for any such homogeneous lattice system.

Let $P_\omega=\chi_{(-\infty,0]}(H_\omega)$ be the projector onto the occupied electron states. Since both the magnetic field and the disorder break the translational invariance, the classical Brillouin torus has no meaning and, understandably, the formula of Eq.~\ref{Chern1} is no longer of any use. However, one can still compute the formula of Eq.~\ref{Chern2}. Thus, a natural extension of the $n${\scriptsize th}-Chern number formula to the class of aperiodic systems is given by:
\begin{equation}\label{Chern3}
C_{n}=\nicefrac{(2 \pi \imath)^n}{n!} \int_\Omega d\mu(\omega)\sum_{\sigma}(-1)^\sigma \mathrm{Tr}\big \{ P_\omega \prod_{i=1}^{2n} \big (\imath[X_{\sigma_i},P_\omega]\big )\chi_{\bm 0} \big \},
\end{equation}
where an average over disorder was also included. 

Now, of course, just because we proposed a formula, one cannot expect that the properties of the Chern numbers, established for the periodic systems, to automatically transfer to the aperiodic case. Nevertheless, we now can formulate the main question to be addressed by the present work: Under what conditions does the quantity defined in Eq.~\ref{Chern3} display quantization and homotopy invariance? As we already mention, the main interest is in the regime of strong disorder where the spectral gap of $H_0$ is filled with dense localized spectrum.

\section{The non-commutative Brillouin torus and $n${\scriptsize th}-Chern number}\label{NCBT}

The non-commutative Brillouin torus provides a natural theoretical framework for the analysis of aperiodic crystals. For the model in Eq.~\ref{LatticeHamOmega}, it is constructed as it follows \cite{BellissardLN2003bv}. Consider the algebra ${\cal A}_0$ given by the space of continuous functions with compact support:
\begin{equation}
f:\Omega \times \mathbb{Z}^{2n} \rightarrow {\cal M}_{Q \times Q},
\end{equation}
where ${\cal M}_{Q \times Q}$ is the space of $Q \times Q$ complex matrices, together with the algebraic operations:
\begin{equation}\label{AlgRules}
\begin{array}{l}
(f+g)(\omega,{\bm x})=f(\omega,{\bm x})+g(\omega,{\bm x}), \medskip \\
(f*g)(\omega,{\bm x})=\sum\limits_{{\bm y} \in {\mathbb Z}^{2n}} e^{\imath \pi( {\bm y}, \hat{{\bm B}} {\bm x})}f(\omega, {\bm y})g(\mathfrak{t}_{{\bm y}}^{-1}\omega,{\bm x}-{\bm y}).
\end{array}
\end{equation} 
Each element from ${\cal A}_0$ defines a family of covariant bounded operators on $\ell^2(\mathbb{Z}^{2n},\mathbb{C}^Q)$, through the representation:
\begin{equation}\label{OpRep}
(\pi_\omega f)|{\bm x},\alpha \rangle =\sum_{{\bm y} \in {\mathbb Z}^{2n}}\sum_{\beta=1}^Q e^{ \imath \pi ({\bm y}, \hat{{\bm B}}{\bm x})} f_{\beta \alpha}(\mathfrak{t}^{-1}_{\bm y}\omega, {\bm x}-{\bm y})|{\bm y},\beta\rangle.
\end{equation}
The Hamiltonian of Eq.~\ref{LatticeHamOmega} is generated by the element:
\begin{equation}
h_{\alpha \beta}(\omega,{\bm x})=t_{\bm 0,\bm x}^{\alpha \beta} (\omega).
\end{equation}

If one introduces the following norm on ${\cal A}_0$:
\begin{equation}\label{Norm}
\|f\| = \sup_{\omega \in \Omega} \|\pi_\omega f \|,
\end{equation}
and the $\ast$-operation: 
\begin{equation}
f^\ast(\omega,{\bm x})=f(\mathfrak{t}_{\bm x}^{-1}\omega,-{\bm x})^\dagger,
\end{equation}
then the completion of ${\cal A}_0$ under the norm of Eq.~\ref{Norm} becomes a $C^\ast$-algebra, which will be denoted by ${\cal A}$. The non-commutative differential calculus over ${\cal A}$ is defined by:
\begin{enumerate}
\item Integration:
\begin{equation}\label{Trace}
{\cal T}(f)=\int_\Omega d\mu(\omega) \ \mathrm{tr} \{f(\omega,{\bm 0})\}, 
\end{equation}
\item Derivations ($i=1,\ldots,2n$):
\begin{equation}
(\partial_i f)(\omega,{\bm x}) = \imath x^i f(\omega,{\bm x}).
\end{equation}
\end{enumerate}
Then the triplet $({\cal A},{\cal T},\partial)$ defines a non-commutative manifold called the non-commutative Brillouin torus. 

The properties of the derivations and the integration have been discussed on many occasions (see for example \cite{BELLISSARD:1994xj}). Here we only mention a few facts that are absolutely need for the calculations to follow:
\begin{itemize}
\item The integration is cyclic: 
\begin{equation}
\mathcal T(f*g)=\mathcal T(g*f).
\end{equation}
\item There is the following equivalent formula of calculus:
\begin{equation}\label{RuleI}
{\mathcal T}(f*\ldots *g)=\int_\Omega d\mu(\omega) \mathrm{Tr}\big \{ (\pi_\omega f ) \ldots  (\pi_\omega g)\chi_{\bm 0} \big \}.
\end{equation}
\item The operator representation of the non-commutative derivation is:
\begin{equation}
\pi_\omega(\partial_i f)=\imath[X_i,\pi_\omega f].
\end{equation}
\item For $1 \leq s < \infty$, the following equation:
\begin{equation}
\|f\|_{L^s}= {\cal T}\left (\{f*f^*\}^\frac{s}{2}\right)^{\frac{1}{s}}
\end{equation}
defines a norm on ${\cal A}_0$. The completion of ${\cal A}_0$ under this norm is called the non-commutative $L^s$-space and is denoted by $L^s({\cal A},{\cal T})$.
\item Let $\alpha =(\alpha_1,\ldots, \alpha_{2n})$ be a multi-index, $|\alpha|=\alpha_1+\ldots \alpha_{2n}$, and $\partial^\alpha=\partial_1^{\alpha_1} \ldots \partial_{2n}^{\alpha_{2n}}$. Then, for $1 \leq s < \infty$ and $k$ a positive integer, the following equation:
\begin{equation}
\|f\|_{W^{s,k}}=\sum_{0\leq |\alpha| \leq k} \|\partial^\alpha f\|_{L^s}
\end{equation}
defines a norm on ${\cal A}_0$. The completion of ${\cal A}_0$ under this norm is called the non-commutative Sobolev space and is denoted by $W^{s,k}({\cal A},{\cal T})$. The non-commutative Sobolev space  $W^{2n,1}({\cal A},{\cal T})$ will play a special role in what follows. Since this is the only Sobolev space used in our work, we will use the simplified notation $W({\cal A},{\cal T})$ for it. It will be useful to explicitly write its norm:
\begin{equation}\label{SobolevNorm}
\|f\|_{W}= {\cal T}\left (|f|^{2n}\right )^{\frac{1}{2n}}+\sum_{i=1}^{2n} {\cal T}\left ( |\partial_i f|^{2n}\right )^{\frac{1}{2n}},
\end{equation}
where $|f|=(f*f^*)^{\frac{1}{2}}$.
\end{itemize} 

Now, let $p=\chi_{(-\infty,0]}(h)$ be the element which generates the covariant family of the spectral projectors $P_\omega$ ($=\pi_\omega p$). Assuming that $p\in W({\cal A},{\cal T})$ (so that everything is finite in Eq.~\ref{Chern4}), then the Chern number formula of Eq.~\ref{Chern3} can be automatically translated over the non-commutative Brillouin torus:
\begin{equation}\label{Chern4}
C_{n}=\nicefrac{(2 \pi \imath)^n}{n!} \sum_{\sigma}(-1)^\sigma {\mathcal T} \big ( p \prod_{i=1}^{2n} \partial_{\sigma_i}p\big ).
\end{equation}
We will refer to the quantity defined in Eq.~\ref{Chern4} as the non-commutative $n${\scriptsize th}-Chern number. The equivalence between this formula and the one in Eq.~\ref{Chern3} can be easily established using the representations $\pi_\omega$ defined in Eq.~\ref{OpRep}.

\section{Quantization and homotopy invariance of $C_n$ at strong disorder}

\subsection{Construction of the Fredholm module}

Let $\gamma_1, \ldots, \gamma_{2n}$ be an irreducible representation of the Clifford algebra $Cl_{2n,0}$ in the finite $2^n$-dimensional Hilbert space $\mathrm{Cliff}(2n)$:
\begin{equation}
\gamma_i \gamma_j + \gamma_j \gamma_i = 2 \delta_{ij},
\end{equation}
and:
\begin{equation}
\gamma_0=-\imath^{-n}\gamma_1 \ldots \gamma_{2n}.
\end{equation}
We denote by $\mathrm{tr}_\gamma$ the ordinary trace over $\mathrm{Cliff}(2n)$. Let
\begin{equation}
{\mathcal H}=\ell^2(\mathbb{Z}^{2n},\mathbb{C}^Q)\otimes \mathrm{Cliff}(2n).
\end{equation}
The $C^*$-algebra ${\cal A}$ can be represented on ${\mathcal H}$ by $\pi_\omega \otimes \mathrm{id}$, $\omega\in \Omega$. We will denote these representations by the same symbols $\pi_\omega$. It is clear that $\pi_\omega(f) \gamma_0=\gamma_0 \pi_\omega(f)$. We will also use the same symbol $\mathrm{Tr}$ for the trace over the Hilbert space ${\cal H}$. 

Let us now define the Dirac operator:
\begin{equation}
D=\sum_{i=1}^{2n} X^i \gamma_i,
\end{equation}
acting by multiplication on ${\mathcal H}$. Throughout our work, we will use the shorthands ${\bm v} \cdot {\bm \gamma}=\sum_{i=1}^{2n} v^i \gamma_i $ and $\hat{{\bm v}}={\bm v}/|{\bm v}|$. Also,
\begin{equation}
D_{\bm a}=({\bm X}+{\bm a})\cdot {\bm \gamma}
\end{equation}
will denote the translated Dirac operator. 

Now, let ${\bm x}_0$ be a fixed point in $\mathbb{R}^{2n}$ such that $0 \leq x_0^i \leq 1$, $i=1,\ldots,2n$. If ${\bm x}_0\notin \mathbb{Z}^{2n}$, we define:
\begin{equation}\label{F1}
\hat{D}_{{\bm x}_0}=\frac{D_{{\bm x}_0}}{|D_{{\bm x}_0}|},
\end{equation}
which acts on ${\cal H}$ by multiplication with $\widehat{{\bm x}+{\bm x}_0}\cdot {\bm \gamma}$.  If ${\bm x}_0\in \mathbb{Z}^{2n}$, we define:
\begin{equation}\label{F2}
(\hat{D}_{{\bm x}_0}\psi)(\bm x)=\left \{
\begin{array}{l}
(\widehat{{\bm x}+{\bm x}_0}\cdot {\bm \gamma})\psi(\bm x), \ \mbox{if} \ {\bm x} \neq -{\bm x}_0 \medskip \\
 ( \frac{1}{\sqrt{2n}}\sum_{i=1}^{2n}\gamma_i ) \psi(\bm x) \ \mbox{if} \ {\bm x} = -{\bm x}_0.
\end{array}
\right .
\end{equation}
Clearly, for all cases, $\hat D_{{\bm x}_0}$ has the following properties: 
\begin{equation}\label{DId}
(\hat{D}_{{\bm x}_0})^2=1, \ \hat{D}_{{\bm x}_0}\gamma_0=-\gamma_0 \hat{D}_{{\bm x}_0}.
\end{equation}
In addition, one can easily show that, for any $f\in \mathcal A$, the operator $i[\hat{D}_{{\bm x}_0},\pi_\omega(f)]$ is compact. Summing up all the facts, we have demonstrated:

\medskip \noindent {\bf Proposition 1.} The triple, $({\mathcal H},\hat{D}_{{\bm x}_0},\gamma_0)$ is an even Fredholm module over ${\mathcal A}$, as defined by Connes \cite{Connes:1994wk}. \medskip

For $n=1$, this Fredholm module is identical to the one used in the work \cite{BELLISSARD:1994xj} on the Integer Quantum Hall Effect.

\subsection{The main statement and its proof}

\noindent{\bf Theorem 2.} Consider the settings and the notations introduced so far. Let $\pi_\omega^\pm(p)$ be the decomposition of $\pi_\omega(p)$ according to the grading induced by $\gamma_0$ on $\mathcal H$. If:
\begin{equation}\label{LocLength}
\Lambda_n=\sum_{i=1}^{2n} {\mathcal T}\left( |\partial_i p|^{2n}\right)^{\frac{1}{2n}}<\infty,
\end{equation}
then, with probability one in $\omega$, the operator $\pi_\omega^-(p)\hat{D}_{{\bm x}_0} \pi_\omega^+(p)$ is in the Fredholm class. Its Fredholm index is independent of $\omega$ or ${\bm x}_0$, and is equal to the non-commutative $n${\scriptsize th}-Chern number, as defined in Eq.~\ref{Chern4}:
\begin{equation}\label{FredholmIndex}
\boxed{
\mathrm{Index}\left(\pi_\omega^-(p)\hat{D}_{{\bm x}_0} \pi_\omega^+(p)\right)=\frac{(2 \pi \imath)^n}{n!} \sum_{\sigma}(-1)^\sigma {\mathcal T} \left ( p \prod_{i=1}^{2n} \partial_{\sigma_i}p\right ).}
\end{equation}
As a consequence, the non-commutative $n${\scriptsize th}-Chern number stays quantized and constant under continuous homotopies of $p$, where the continuity is considered with respect to the Sobolev norm $\| \ \|_W$ of $W({\cal A},{\cal T})$.\medskip

\noindent {\bf Remark.} $\Lambda_n$ has the unit of length and can be interpreted as the natural definition of the localization length for the present problem.

\medskip \noindent{\it Proof.} The conditions of the Theorem together with Lemma~4 (see below), assure that, with probability one in $\omega$, the operator $[\hat{D}_{{\bm x}_0},\pi_\omega(p)]$ belongs to the $q$th-Schatten class, for any $q>2n$. This in turn, implies that the operators:
\begin{equation}
\pi_\omega^\mp(p)-\pi_\omega^\mp(p)\hat{D}_{{\bm x}_0} \pi_\omega^\pm(p)\hat{D}_{{\bm x}_0}\pi_\omega^\mp(p)=\pi_\omega^\mp(p)[\hat{D}_{{\bm x}_0},\pi_\omega(p)]^2\pi_\omega^\mp(p)
\end{equation}
are in the $q$th-Schatten class for any $q>n$. As such, $\pi_\omega^-(p)\hat{D}_{{\bm x}_0} \pi_\omega^+(p)$ belongs to the Fredholm class and its Fredholm index is well defined for all $\omega\in \Omega$, except for a possible zero-measure subset.

Next, we establish three key properties of the Fredholm index of $\pi_\omega^-(p)\hat{D}_{{\bm x}_0} \pi_\omega^+(p)$: 1) it is, with probability one, independent of $\omega$, 2) it is independent of ${\bm x}_0$, and 3) it is independent of what exactly is inserted at the second line of Eq.~\ref{F2}. Indeed, for 1), since the translations act ergodically on $\Omega$, it is enough to investigate what happens when we replace $\omega$ with an arbitrarily translated $\mathfrak{t}_{\bm a}\omega$. And since the Fredholm index is invariant to unitary transformations, and due to the covariance of $\pi_\omega$, we only need to compare the indices of $\pi_\omega^-(p)\hat{D}_{{\bm x}_0} \pi_\omega^+(p)$ and $\pi_\omega^-(p)\hat{D}_{{\bm x}_0+{\bm a}} \pi_\omega^+(p)$. But these two operators differ by a compact operator so their Fredholm indices coincide. Property 2) follows from the same arguments and 3) is evident.

\medskip We now start the actual computation of the Fredholm index. We will use  Connes' Chern characters \cite{Connes:1994wk} for the Fredholm module of Proposition 1. Given that $[\pi_\omega(p),\hat{D}_{{\bm x}_0}]$ is in the $q$th-Schatten class with $q>2n$,  the lowest Chern character we can use is $\tau_{2n}$:
\begin{equation}
\mathrm{Index}\left(\pi_\omega^-(p)\hat{D}_{{\bm x}_0} \pi_\omega^+(p)\right)=\tau_{2n}\big (\pi_\omega(p),\ldots,\pi_\omega(p)\big ),
\end{equation}
where
\begin{equation}
\tau_{2n}\big (\pi_\omega(p),\ldots,\pi_\omega(p)\big )= \mathrm{Tr}_S\{\gamma_0 [\hat{D}_{{\bm x}_0},\pi_\omega(p)]^{2n}\},
\end{equation}
with $\mathrm{Tr}_S$ being the super-trace for even Fredholm modules \cite{Connes:1994wk}. For our particular case:
\begin{eqnarray}\label{QQQ}
\tau_{2n}\big (\pi_\omega(p),\ldots,\pi_\omega(p)\big)=\nicefrac{1}{2}\mathrm{Tr}\left \{\gamma_0[\hat{D}_{{\bm x}_0},\pi_\omega(p)]^{2n+1}\hat{D}_{{\bm x}_0}\right \}.
\end{eqnarray}
The righthand side can be processed to the following form:
\begin{eqnarray}\label{Impo}
-\int_{\mathbb{R}^{2n}}d{\bm x} \int_\Omega d\mu(\omega) \   \mathrm{Tr} \{ \gamma_0 \pi_\omega(p)[\hat{D}_{\bm x},\pi_\omega(p)]^{2n}\chi_{\bm 0} \},
\end{eqnarray}
where it is understood that the equality holds for all ${\bm x}_0$ and all $\omega \in \Omega$, with possible exceptions that occur with zero probability. Indeed, since, with probability one, the index is independent of $\omega$, we can average over this variable. By inserting the resolution of identity $\sum_{\bm x \in \mathbb{Z}^{2n}} \chi_n$ in Eq.~\ref{QQQ}, then expanding and using the translations, we arrive at:
\begin{equation}
\ldots = \nicefrac{1}{2} \int_\Omega d\mu(\omega) \sum_{{\bm x}\in \mathbb{Z}^{2n}}  \mathrm{Tr} \left \{ \gamma_0  [\hat{D}_{{\bm x}+{\bm x}_0},\pi_{\mathfrak{t}_{x}\omega}(p)]^{2n+1}\hat{D}_{{\bm x}+{\bm x}_0} \chi_{\bm 0} \right \}. 
\end{equation}
Since the sum over ${\bm x}$ is absolutely convergent, we can move the integration inside the sums. We then perform a change of variable $\omega\rightarrow {\mathfrak{t}_{x}\omega}$ and use the invariance of the measure $\mu(\omega)$ to write:
\begin{equation}
\ldots = \nicefrac{1}{2} \sum_{{\bm x}\in {\bm x}_0+\mathbb{Z}^{2n}} \int_\Omega d\mu(\omega) \mathrm{Tr} \left \{ \gamma_0 [\hat{D}_{\bm x},\pi_\omega(p)]^{2n+1}\hat{D}_{\bm x} \chi_{\bm 0} \right \}.
\end{equation}
Since the index is independent of ${\bm x}_0$, we can integrate ${\bm x}_0$ over the unit cell and transform the discrete sum over ${\bm x}$ into an integral over the whole space. Next, we open one commutator and use the cyclic properties of the trace and the identities 
in Eqs.~\ref{DId} and
\begin{equation}\label{ids}
\left [ \hat{D}_{{\bm x}_0},[\hat{D}_{{\bm x}_0},\pi_\omega(p)]^2 \right ]  =0,\  \left [\pi_\omega(p),[\hat{D}_{{\bm x}_0},\pi_\omega(p)]^2 \right ]=0,
\end{equation}
 to arrive at Eq.~\ref{Impo}.

We continue the calculation from Eq.~\ref{Impo}. Let $\tilde p \in {\cal A}_0$ be an approximation of $p$ (hence $\tilde{p}$ has compact support). Since ${\cal A}_0$ is dense in the Soboleev space, we can always find a sequence of such approximations that converges to $p$ in the Sobolev norm $\| \ \|_W$. We will evaluate the righthand side of  Eq.~\ref{Impo} with $p$ replaced by $\tilde p$. Eq.~\ref{Impo} can be expended to:
 \begin{equation}
-\int_{\mathbb{R}^{2n}}d{\bm x} \int_\Omega d\mu(\omega) \sum_{{\bm x_i}'s\in \mathbb{Z}^{2n}} \mathrm{Tr} \big \{\gamma_0  \pi_\omega(\tilde p)\prod_{i=1}^{2n} \chi_{{\bm x}_{i}}[\hat{D}_{\bm x},\pi_\omega(\tilde p)] \chi_{{\bm x}_{i+1}} \big \},
\end{equation}
 where ${\bm x}_{2n+1}={\bm 0}$. One important observation is that, due to the compact support of $\tilde p$, the sums over ${\bm x}_i$'s involve a finite number of zero elements. As such, we can interchange the sums and the integrals and, after the commutators are evaluated explictely, we obtain:
\begin{eqnarray}\label{Last1}
\ldots = -\sum_{{\bm x_i}'s\in \mathbb{Z}^{2n}} \int_{\mathbb{R}^{2n}} d{\bm x} \ \mathrm{tr}_\gamma \big \{\gamma_0 \prod_{i=1}^{2n}\left(\widehat{{\bm x}_i+{\bm x}} -\widehat{{\bm x}_{i+1}+{\bm x}}\right )\cdot  {\bm \gamma}\big \} \nonumber \\
 \indent \indent \indent \times \int_\Omega d\mu(\omega) \mathrm{Tr}\big \{\pi_\omega(\tilde p) \prod_{i=1}^{2n} \chi_{{\bm x}_{i}}\pi_\omega(\tilde p) \chi_{{\bm x}_{i+1}} \big \},
 \end{eqnarray}
Using the key identity from Lemma~3, this is the same as:
 \begin{equation}
 \ldots =  \nicefrac{(-2\pi \imath)^n}{n!} \sum_{{\bm x_i}'s\in \mathbb{Z}^{2n}} \sum_\sigma (-1)^\sigma \prod_{i=1}^{2n} \left (x_i^{\sigma_i} \right ) 
 \int_\Omega d\mu(\omega)\mathrm{Tr} \big\{ \pi_\omega(\tilde p) \prod_{i=1}^{2n} \chi_{{\bm x}_{i}}\pi_\omega(\tilde p) \chi_{{\bm x}_{i+1}} \big \}, \nonumber
 \end{equation}
 where $\sigma$'s denote permutations of ${1,\ldots,2n}$. This can be conveniently written as:
\begin{equation}
  \ldots = \nicefrac{(-2\pi \imath)^n}{n!}\int_\Omega d\mu(\omega) \sum_\sigma (-1)^\sigma \mathrm{Tr}\big \{ \pi_\omega(\tilde p)\prod_{i=1}^{2n} [X_{\sigma_i},\pi_\omega(\tilde p)] \chi_{\bm 0}\big \},
  \end{equation}
  and using the rule of calculus from Eq.~\ref{RuleI}, we finally obtain:
  \begin{equation}
  \ldots = \nicefrac{(2\pi \imath)^n}{n!}\sum_\sigma (-1)^\sigma {\cal T}\left (\tilde p \prod_{i=1}^{2n} \partial_{\sigma_i}\tilde p \right ).
  \end{equation}
We now can take the limit of $\tilde p$ towards $p$ in the Sobolev space and the statement follows. Regarding the homotopy invariance, we only need to observe that if $p$ is varied continuously with respect to the Sobolev norm $\| \ \|_W$, then the right-hand side of Eq.~\ref{FredholmIndex} varies continuously and, as a consequence, it cannot jump from one quantized value to another. \qed

\medskip \noindent{\bf Lemma 3.} Let ${\bm x}_1,\ldots,{\bm x}_{2n+1}$ be points of $\mathbb{R}^{2n}$ with ${\bm x}_{2n+1}={\bm 0}$. Then the following identity holds:
\begin{eqnarray}\label{Identity}
\boxed{
\int\limits_{\mathbb{R}^{2n}}d{\bm x} \ \mathrm{tr}_\gamma \big \{\gamma_0 \prod_{i=1}^{2n}\left(\widehat{{\bm x}_i+{\bm x}} -\widehat{{\bm x}_{i+1}+{\bm x}} \right )\cdot  {\bm \gamma}\big \} 
 =-\frac{(2\pi )^n}{\imath^n n!}\sum_{\sigma} (-1)^\sigma \prod_{i=1}^{2n} x_i^{\sigma_i}.}
\end{eqnarray} 

\medskip \noindent {\it Proof.} The proof of this key identity relies on a geometric interpretation of the following trace:
\begin{equation}\label{GeomId}
\mathrm{tr}_\gamma\{\gamma_0 ({\bm y}_1 \cdot {\bm \gamma}) \ldots ({\bm y}_{2n} \cdot {\bm \gamma})\}=-\imath^{-n} 2^{n} n! \  \mathrm{Vol}[{\bm 0},{\bm y}_1,\ldots,{\bm y}_{2n}],
\end{equation}
where $[{\bm y}_0,{\bm y}_1,\ldots,{\bm y}_{2n}]$ is the simplex with vertices at ${\bm y}_0$, ${\bm y}_1$, $\ldots$, ${\bm y}_{2n}$. $\mathrm{Vol}$ denotes the oriented volume of the simplex. Note that ${\bm y}_0={\bm 0}$ in Eq.~\ref{GeomId}. The identity in Eq.~\ref{GeomId} follows by observing that:
\begin{equation}\label{bb}
\mathrm{tr}_\gamma\{\gamma_0 ({\bm y}_1 \cdot {\bm \gamma}) \ldots ({\bm y}_{2n} \cdot {\bm \gamma})\}=-\imath^{-n} 2^{n} \det({\bm y}_1,\ldots,{\bm y}_{2n}),
\end{equation}
where inside the determinant we have the $2n \times 2n$ matrix with ${\bm y}_{i}$ as columns. The $2^{n}$ factor in front appears when tracing the identity matrix on $\mathrm{Cliff}(2n)$ and the factor $-\imath^{-n}$ comes from the definition of $\gamma_0$. According to Ref.~\cite{SteinAMM1966bv}, the determinant in Eq.~\ref{bb} is nothing but $n!$ times the volume of the simplex $[{\bm 0},{\bm y}_1,\ldots,{\bm y}_{2n}]$.

Expanding the lefthand side of Eq.~\ref{Identity}, we obtain:
\begin{equation}\label{X1}
\ldots = -\imath^{-n} 2^{n} n! \int_{\mathbb{R}^{2n}}d{\bm x}\sum_{j=1}^{2n+1}(-1)^{j+1} \mathrm{Vol}[{\bm 0},\widehat{{\bm x}_1-{\bm x}},\ldots, \widehat{\underline{{\bm x}_j-{\bm x}}},\ldots,\widehat{{\bm x}_{2n+1}-{\bm x}}],
\end{equation}
where the underline means the term is omitted. In Eq.~\ref{X1}, it is convenient to translate the simplexes and move the first vertex to the proper place. As such, we will work with:
\begin{equation}
\mathfrak{S}_j({\bm x})=[{\bm x}+\widehat{{\bm x}_1-{\bm x}},\ldots, {\bm x},\ldots,{\bm x}+\widehat{{\bm x}_{2n+1}-{\bm x}}], \  j=1,\ldots,2n+1,
\end{equation}
where ${\bm x}$ is located at the $j$-th position. We will also denote by $\mathfrak{S}$ the simplex:
\begin{equation}
\mathfrak{S}=[{\bm x}_1, \ldots, {\bm x}_{2n+1}],
\end{equation}
where we recall that ${\bm x}_{2n+1}$ coincides with the origin. To summarize, we arrived at:
\begin{eqnarray}\label{Intermetzo}
\int_{\mathbb{R}^{2n}}d{\bm x} \ \mathrm{tr}_\gamma \big \{\gamma_0 \prod_{i=1}^{2n}\big(\widehat{ {\bm x}_i-{\bm x}}-\widehat{ {\bm x}_{i+1}-{\bm x}}\big )\cdot{\bm \gamma}\big \}  \nonumber \\
\indent \indent \indent  =-n! (-2\imath)^n\int_{\mathbb{R}^{2n}}d{\bm x}\sum_{j=1}^{2n+1} \mathrm{Vol}\{\mathfrak{S}_j({\bm x})\}.
\end{eqnarray}
Note that the sign factor $(-1)^{j+1}$ disappeared because we changed the order of the vertices. Furthermore, if ${\bm x}$ is located inside $\mathfrak{S}$, then the orientations of the simplexes $\mathfrak{S}_j$ ($j=\overline{1,2n+1}$) are the same as that of $\mathfrak{S}$. This is the case because each $\mathfrak{S}_j$ can be continuously deformed into $\mathfrak{S}$ without sending its volume to zero. Such deformation can be achieved by moving ${\bm x}$ at ${\bm x}_j$ and ${\bm x}+\widehat{{\bm x}_i-{\bm x}}$ at ${\bm x}_i$ ($i\neq j$), along straight paths. One such deformation process is illustrated in Fig.~1(a) for the 2-dimensional case.

\begin{figure}
\center
  \includegraphics[width=12cm]{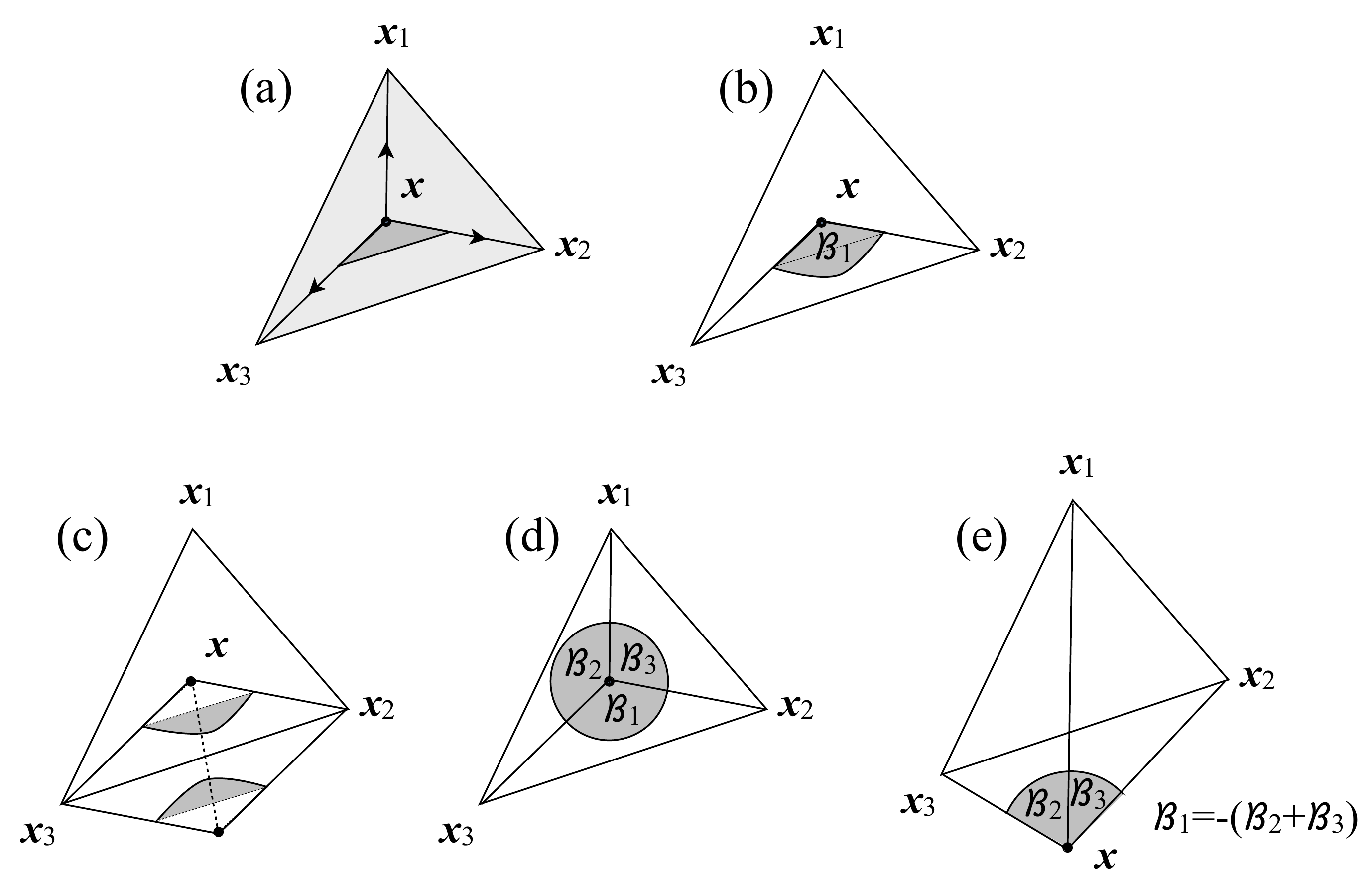}\\
  \caption{(a) Illustration of the simplexes $\mathfrak{S}=({\bm x}_1,{\bm x}_2,{\bm x}_3)$ (light gray) and $\mathfrak{S}_1({\bm x})=({\bm x},{\bm x}+\widehat{{\bm x}_2-{\bm x}},{\bm x}+\widehat{{\bm x}_3-{\bm x}})$ (darker gray), together with the interpolation process that takes $\mathfrak{S}_1({\bm x})$ into $\mathfrak{S}$. (b) Illustration of the ball sector ${\cal B}_1({\bm x})$ corresponding to the simplex $\mathfrak{S}_1({\bm x})$. (c) Illustration of the inversion operation on ${\bm x}$ relative to the center of the segment $({\bm x}_2,{\bm x}_3)$. The volume of ${\cal B}_1({\bm x})-\mathfrak{S}_1({\bm x})$ (shaded in gray) changes sign after this operation because ${\bm x}$ crosses the segment $({\bm x}_2,{\bm x}_3)$ (the volume becomes zero at the crossing and changes sign after that). (d) The ball sectors ${\cal B}_1({\bm x})$, ${\cal B}_2({\bm x})$ and ${\cal B}_3({\bm x})$ have same orientation and they add up to the full unit disk when ${\bm x}$ is inside $\mathfrak{S}$. (e) When ${\bm x}$ is outside of $\mathfrak{S}$, the ball sector ${\cal B}_1({\bm x})$ has the opposite orientation of ${\cal B}_2({\bm x})$ and ${\cal B}_3({\bm x})$ (because ${\bm x}$ crossed the segment $({\bm x}_2,{\bm x}_3)$) and, as a consequence, ${\cal B}_1({\bm x})$, ${\cal B}_2({\bm x})$ and ${\cal B}_3({\bm x})$ add up to zero.}
\end{figure} 

If we look closer at the simplexes $\mathfrak{S}_j({\bm x})$, we see that all vertices that are different from ${\bm x}$ are located on the unit sphere centered at ${\bm x}$. As such, for any given simplex $\mathfrak{S}_j({\bm x})$, the facets stemming from ${\bm x}$ define a sector of the unit ball ${\bm {\mathcal B}}_{2n}({\bm x})$ centered at ${\bm x}$. This sector will be denoted by ${\cal B}_j({\bm x})$. This construction is illustrated in Fig.~1(b) for the 2-dimensional case. The orientation of ${\cal B}_j({\bm x})$'s are considered to be the same as that of $\mathfrak{S}_j({\bm x})$'s. Now, one important observation is that:
\begin{equation}\label{Asymptotic}
\mathrm{Vol}\{\mathfrak{S}_j({\bm x})\}-\mathrm{Vol}\{ {\cal B}_j({\bm x})\} \sim |{\bm x}|^{-(2n+1)}
\end{equation}
in the asymptotic regime $|{\bm x}|\rightarrow \infty$. Consequently, we can write:
\begin{eqnarray}
\int_{\mathbb{R}^{2n}}d{\bm x} \sum_{i=1}^{2n+1} \mathrm{Vol}\{ \mathfrak{S}_j ({\bm x})\}= \int_{\mathbb{R}^{2n}}d{\bm x} \sum_{i=1}^{2n+1} \mathrm{Vol}\{ {\cal B}_j ({\bm x})\}\\
\indent +\sum_{i=1}^{2n+1} \int_{\mathbb{R}^{2n}}d{\bm x} \big ( \mathrm{Vol}\{ \mathfrak{S}_j ({\bm x})\}-\mathrm{Vol}\{ {\cal B}_j ({\bm x}) \} \big ), \nonumber
\end{eqnarray}
with the integrals in the second line being absolutely convergent. There are two extraordinary facts taking place, simultaneously:
\begin{equation}\label{Fact1}
\int_{\mathbb{R}^{2n}}d{\bm x} \big ( \mathrm{Vol}\{ \mathfrak{S}_j ({\bm x})\}-\mathrm{Vol}\{ {\cal B}_j ({\bm x}) \} \big )=0, \ \mbox{for all} \ j=\overline{1,2n+1},
\end{equation}
and
\begin{equation}\label{Fact2}
 \sum_{i=1}^{2n+1} \mathrm{Vol}\{ {\cal B}_j ({\bm x})\}=\left \{
 \begin{array}{l}
 \mathrm{Vol}({\bm {\mathcal B}}_{2n}) \ \mbox{if} \ {\bm x} \ \mbox{inside} \ \mathfrak{S},\medskip \\
 0 \ \mbox{if} \ {\bm x} \ \mbox{outside} \ \mathfrak{S}.
 \end{array}
 \right .
 \end{equation}
Eq.~\ref{Fact1} follows from the fact that the integrand is odd under the inversion of ${\bm x}$ relative to the center of the facet $\{ {\bm x}_1,\ldots,\underline{{\bm x}_j},\ldots,{\bm x}_{2n+1}\}$ of the simplex $\mathfrak{S}$. This property is illustrated in Fig.~1(c) for the 2-dimensional case. Eq.~\ref{Fact2} is a geometric fact. It is illustrated in Figs.~1(d) and 1(e) for the 2-dimensional case. Then Eq.~\ref{Intermetzo} reduces to:
\begin{equation}
\int_{\mathbb{R}^{2n}}d{\bm x} \ \mathrm{tr}_\gamma \big \{\gamma_0 \prod_{i=1}^{2n}\left(\widehat{ {\bm x}_i-{\bm x}}-\widehat{ {\bm x}_{i+1}-{\bm x}}\right )\cdot {\bm \gamma}\big \} = -n! (-2\imath)^n \mathrm{Vol}\{{\bm {\mathcal B}}_{2n})\}\mathrm{Vol}\{ \mathfrak{S} \},
\end{equation}
and the statement follows by writing the volume of the simplex $\mathfrak{S}$ as a determinant like in Ref.~\cite{SteinAMM1966bv}.\qed

\subsection{The Dixmier trace calculation}

\medskip \noindent{\bf Lemma 4.} Let $p$ be an element from the Sobolev space $W({\cal A},{\cal T})$ and let $\mathrm{Tr}_{\mathrm{Dix}}$ denote the Dixmier trace \cite{Dixmier1966jc}. Then, with probability one in $\omega$, the following identity holds:
\begin{equation}\label{DixmierTrace}
\boxed{
\mathrm{Tr}_{\mathrm{Dix}}\left \{\big (\imath[\hat{D}_{{\bm x}_0},\pi_\omega(p)] \big)^{2n}\right \}=\frac{1}{2n}\int\limits_{\mathcal S_{2n-1}}d\hat{{\bm x}} \ \mathcal T \otimes \mathrm{tr}_{\gamma}\left \{ \big({\bm \gamma}(\hat{{\bm x}})\cdot {\bm \nabla p} \big )^{2n}\right \}.}
\end{equation}
In particular, the Dixmier trace of $(i[\hat{D}_{{\bm x}_0},\pi_\omega(p)])^{2n}$ is finite for any $p\in W({\cal A},{\cal T})$, a fact that follows immediately from the non-commutative version of the Holder inequality \cite{SegalAM1953bv,KosakiJFA1984gf}.
In Eq.~\ref{DixmierTrace}, $\hat{{\bm x}}$ is the unit vector in $\mathbb{R}^{2n}$ which is integrated over the (2$n$-1)-sphere ${\cal S}_{2n-1}$, and 
\begin{equation}
\gamma_i(\hat{{\bm x}})=\gamma_i -\hat x_i (\hat{\bm x}\cdot {\bm \gamma}), \ i=1,\ldots,2n,
\end{equation}
 are the generators of the Clifford algebra:
\begin{equation}
\{\gamma_i(\hat{{\bm x}}),\gamma_j(\hat{{\bm x}})\}=2(\delta_{ij}-\hat{x}_i \hat{x}_j).
\end{equation}

\medskip \noindent{\bf Remark.} From a standard property of the Dixmier trace \cite{BELLISSARD:1994xj}, it follows that $\imath[\hat{D}_{{\bm x}_0},\pi_\omega(p)] $ belongs to any $q$th-Schatten classe with $q>2n$, whenever $p\in W({\cal A},{\cal T})$.

\medskip \noindent {\it Proof.} We will derive the identity in Eq.~\ref{DixmierTrace} for an element $p$ from ${\cal A}_0$ (hence with compact support). Since ${\cal A}_0$ is dense in the Sobolev space $W({\cal A},{\cal T})$, the identity extends by continuity over this space. First, it is easy to establish that any translation of $\omega$ leads to a trace-class perturbation and, as such, the Dixmier trace remains unchanged. Since the translations act ergodically on $\Omega$, this tells that the Dixmier trace in Eq.~\ref{DixmierTrace} is independent of $\omega$, except for cases that occur with zero probability. Using the arguments from Ref.~\cite{BELLISSARD:1994xj}, we only need to consider the diagonal part of the operator:
\begin{equation}
\mathrm{Diag} \ \big (\imath[\hat{D},\pi_\omega(p)] \big)^{2n}=(-1)^n\sum_{{\bm x}\in \mathbb{Z}^{2n}}\sum_{\alpha=1}^Q\mathrm{tr}_\gamma \left \{ \langle {\bm x},\alpha |[\hat{D},\pi_\omega(p)]^{2n}|{\bm x},\alpha \rangle \right \} \chi_{{\bm x},\alpha},
\end{equation}
where $\chi_{{\bm x},\alpha}=|{\bm x},\alpha \rangle \langle {\bm x},\alpha |$. We expand the righthand side in the following way:
\begin{equation}
 \dots = (-1)^n\sum_{{\bm x}\in \mathbb{Z}^{2n}}\sum_{\alpha=1}^Q \sum_{{\bm x}_i's \in \mathbb{Z}^{2n}}\mathrm{tr}_\gamma \big \{ \langle {\bm 0},\alpha |\prod_{i=1}^{2n}\chi_{{\bm x}_{i-1}}[\hat{D}_{\bm x},\pi_{\mathfrak{t}_{\bm x}\omega}(p)]\chi_{{\bm x}_{i}}|{\bm 0},\alpha \rangle \big \} \chi_{{\bm x},\alpha}, 
\end{equation}
where ${\bm x}_0={\bm x}_{2n}={\bm 0}$ and the summation over these variables is omitted. The commutators can be evaluated explicitly and note that, since $p$ has a compact support, there are only a finite number of non-zero terms in the sums over ${\bm x}_i$ variables. We have:
\begin{eqnarray}
\ldots = (-1)^n \sum_{{\bm x}\in \mathbb{Z}^{2n}}\sum_{\alpha=1}^Q \sum_{{\bm x}_i's \in \mathbb{Z}^{2n}}\langle {\bm 0},\alpha |\prod_{i=1}^{2n}\chi_{{\bm x}_{i-1}}\pi_{\mathfrak{t}_{\bm x}\omega}(p)\chi_{{\bm x}_{i}}|{\bm 0},\alpha \rangle \\
\indent \indent \indent \times \mathrm{tr}_\gamma \big \{\prod_{i=1}^{2n}  \left ( \widehat{{\bm x}_{i-1}+{\bm x}} - \widehat{{\bm x}_{i}+{\bm x}} \right )\cdot{\bm \gamma}  \big \}\chi_{{\bm x},\alpha}. \nonumber
\end{eqnarray}
In the asymptotic limit $|{\bm x}|\rightarrow \infty$, we have:
\begin{equation}
  \left ( \widehat{{\bm x}_{i-1}+{\bm x}} - \widehat{{\bm x}_{i}+{\bm x}} \right )\cdot{\bm \gamma}  
=\frac{1}{|{\bm x}|}  ({\bm x}_{i-1}-{\bm x_i})\cdot{\bm \gamma}( \hat{{\bm x}}\big ) +O(|{\bm x}|^{-2}).
\end{equation}
Hence, apart from terms that are in the trace-class and don't count for the Dixmier trace \cite{BELLISSARD:1994xj}, the diagonal part of our operator is:
\begin{equation}
\mathrm{Diag}\ \big (\imath[\hat{D},\pi_\omega(p)] \big)^{2n} = \sum_{\alpha=1}^Q \sum_{{\bm x}\in \mathbb{Z}^{2n}} \mathrm{tr}_\gamma \big \{\langle {\bm 0},\alpha |\big ({\bm \gamma}(\hat{\bm x}) \cdot \imath[{\bm X},\pi_{\mathfrak{t}_{\bm x}\omega}(p)] \big)^{2n}|{\bm 0},\alpha \rangle \big \}  \frac{\chi_{{\bm x},\alpha}}{|{\bm x}|^{2n}}.
\end{equation}
Then the statement follows from the following Lemma.

\medskip \noindent {\bf Lemma 5.} Let $f:\Omega \rightarrow \mathbb{C}$ and $\varphi: {\cal S}_{2n-1}\rightarrow \mathbb{C}$ be bounded measurable functions. Consider the operator:
\begin{equation}\label{Ope1}
\sum_{{\bm x}\in \mathbb{Z}^{2n}}  f(\mathfrak{t}_{\bm x}\omega)\varphi(\hat{\bm x})\frac{\chi_{{\bm x},\alpha}}{|{\bm x}|^{2n}}, 
\end{equation}
and assume that any translation of $\omega$ in Eq.~\ref{Ope1} leads to a trace-class perturbation. Then:
\begin{equation}\label{Dix}
\boxed{
\mathrm{Tr}_{\mathrm{Dix}} \sum_{{\bm x}\in \mathbb{Z}^{2n}}  f(\mathfrak{t}_{\bm x}\omega)\varphi(\hat{\bm x})\frac{\chi_{{\bm x},\alpha}}{|{\bm x}|^{2n}} 
=\frac{1}{2n}\int\limits_\Omega d\mu(\omega) \ f(\omega)\int\limits_{{\mathcal S}_{2n-1}} d \hat{\bm x} \ \varphi(\hat{\bm x}).}
\end{equation}

\medskip \noindent {\it Proof.} One of the main ingredients of the proof is Lemma 3 of Ref.~\cite{BELLISSARD:1994xj}, which says:
\begin{equation}
\mathrm{Tr}_{\mathrm{Dix}} \sum_{{\bm x}\in \Sigma} \frac{\chi_{{\bm x},\alpha}}{|{\bm x}|^{2n}}=\frac{s_{2n-1}}{2n}\mathrm{Dens} \ \Sigma,
\end{equation}
where $s_{2n-1}$ is the area of the (2$n$-1)-sphere, and the density of a set $\Sigma \in \mathbb{Z}^{2n}$ is defined as:
\begin{equation}
\mathrm{Dens} \ \Sigma=\lim_{N\rightarrow \infty} \frac{1}{N^{2n}}\sum_{{\bm x}\in \Sigma \cap \mathfrak{C}_{2n}} 1,
\end{equation}
with $\mathfrak{C}_{2n}$ being the cube $\left [-\frac{N}{2},\frac{N}{2}\right ]^{2n}$. To use this general result, we partition the space according to the level sets of $\varphi$ and $f$. First, let us partition the configuration space $\Omega$:
\begin{equation}
\Omega_j=\{\omega\in \Omega \ | \  f_{j-1} \leq f(\omega) <f_j \},
\end{equation}
where $f_j=( j+\nicefrac{1}{2})\delta$ and $\delta$ is a small positive number. Since $f$ is bounded, only a finite number of $\Omega_j$'s are non-empty. Then let $\Sigma_j(\omega)$ be the sets in $\mathbb{Z}^{2n}$, defined as:
\begin{equation}
\Sigma_j(\omega)=\{{\bm x}\in \mathbb{Z}^{2n}\  | \  \mathfrak{t}_x \omega \in \Omega_j \}.
\end{equation}
These sets have finite densities in $\mathbb{Z}^{2n}$, and in fact Birckhoff's theorem tells us that, with probability one:
\begin{equation}
\mathrm{Dens} \ \Sigma_j(\omega)=\mu(\Omega_j).
\end{equation} 

The unit vector $\hat{\bm x}$ takes values on the ${\cal S}_{2n-1}$ unit sphere in $\mathbb{R}^{2n}$, and we partition the sphere into level sets of $\varphi$:
\begin{equation}
S_k=\{ \hat{\bm x}\in {\cal S}_{2n-1}\ | \ \varphi_{k-1} \leq \varphi(\hat{\bm x}) < \varphi_k  \},
\end{equation}
where $\varphi_k=(k+\nicefrac{1}{2})\delta$. Since $\varphi$ is bounded, there are a finite number of non-empty such sets. We then refine our partition of $\mathbb{Z}^{2n}$ to:
\begin{equation}
\Sigma_{j,k}(\omega)=\Sigma_j(\omega) \cap \{ {\bm x}\in \mathbb{R}^{2n} \ | \ \hat{\bm x}\in  S_k \}.
\end{equation} 
This partition is still finite, hence we can write:
\begin{equation}
\mathrm{Tr}_{\mathrm{Dix}} \sum_{{\bm x}\in \mathbb{Z}^{2n}} f(\mathfrak{t}_{\bm x}\omega)\varphi(\hat{\bm x})\frac{\chi_{{\bm x},\alpha}}{|{\bm x}|^{2n}}  =\sum_{j,k}\mathrm{Tr}_{\mathrm{Dix}} \sum_{{\bm x}\in \Sigma_{j,k}(\omega)} f(\mathfrak{t}_{\bm x}\omega)\varphi(\hat{\bm x})\frac{\chi_{{\bm x},\alpha}}{|{\bm x}|^{2n}}.
\end{equation}
Furthermore, from the definition of the partition:
\begin{equation}\label{Rel1}
\mathrm{Tr}_{\mathrm{Dix}} \sum_{{\bm x}\in \Sigma_{j,k}(\omega)} f(\mathfrak{t}_{\bm x}\omega)\varphi(\hat{\bm x})\frac{\chi_{{\bm x},\alpha}}{|{\bm x}|^{2n}} =
f_j \phi_k \ \mathrm{Tr}_{\mathrm{Dix}} \sum_{{\bm x}\in \Sigma_{j,k}(\omega)} \frac{\chi_{{\bm x},\alpha}}{|{\bm x}|^{2n}},
\end{equation}
plus corrections that are of second order in $\delta$. We arrive at: 
\begin{equation}\label{Rel2}
\mathrm{Tr}_{\mathrm{Dix}} \sum_{{\bm x}\in \Sigma_{j,k}(\omega)} f(\mathfrak{t}_{\bm x}\omega)\varphi(\hat{\bm x})\frac{\chi_{{\bm x},\alpha}}{|{\bm x}|^{2n}} =
 f_j \phi_k \ \frac{s_{2n-1}}{2n}\mathrm{Dens} \ \Sigma_{j,k}(\omega),
\end{equation}
plus the corrections we mentioned. Our next task will be to compute the density of $\Sigma_{j,k}(\omega)$. However, instead of working directly with $\Sigma_{j,k}(\omega)$, we recall that the original Dixmier trace remains unchanged when we translate $\omega$. As such, we can replace $\mathrm{Dens} \ \Sigma_{j,k}(\omega)$ with the average $M^{-2n}\sum_{{\bm m}\in \mathfrak{C}_M}\mathrm{Dens} \ \Sigma_{j,k}(\mathfrak{t}_{\bm m}\omega)$. The latter can also be written as:
\begin{equation}
M^{-2n}\sum_{{\bm m}\in \mathfrak{C}_M} \mathrm{Dens} \left \{ [{\bm m}+\Sigma_{j}(\omega)]\cap \{ {\bm x}\in \mathbb{R}^{2n} \ | \ \hat{\bm x}\in S_k \} \right \}.
\end{equation}
We will work in the limit $M \rightarrow \infty$, where we consider the following measure on the sphere:
\begin{equation}
\nu(S\subset {\mathcal S}_{2n-1})=\lim_{M\rightarrow \infty}\frac{M^{-2n}}{\mu(\Omega_j)}\sum_{{\bm m}\in \mathfrak{C}_M} \mathrm{Dens} \left \{ [{\bm m}+\Sigma_{j}(\omega)]\cap \{ {\bm x}\in \mathbb{R}^{2n} \ | \ \hat{\bm x}\in S \} \right \}.
\end{equation}
From this expression, one can derive two immediate properties. First, $\nu({\mathcal S}_{2n-1})=1$, hence $\nu$ is a probability measure, and second, $\nu$ is invariant to rotations. Hence $\nu$ must be equal to:
\begin{equation}
\nu(S\subset {\mathcal S}_{2n-1})=\frac{|S|}{s_{2n-1}},
\end{equation} 
where $|S|$ denotes the area of $S$. Consequently:
\begin{equation}
\lim_{M\rightarrow 0} M^{-2n}\sum_{{\bm m}\in \mathfrak{C}_M}\mathrm{Dens} \ \Sigma_{j,k}(\mathfrak{t}_{\bm m}\omega) = \frac{\mu(\Omega_j)|S|}{s_{2n-1}}.
\end{equation}
Putting everything together, we demonstrated that, apart from corrections that vanish as $\delta \rightarrow 0$:
\begin{equation}
\mathrm{Tr}_{\mathrm{Dix}} \sum_{{\bm x}\in \mathbb{Z}^{2n}} f(\mathfrak{t}_{\bm x}\omega)\varphi(\hat{\bm x})\frac{\chi_{{\bm x},\alpha}}{|{\bm x}|^{2n}}  =\frac{1}{2n}\sum_{j,k} f_j \phi_k \mu(\Omega_j)|S_k|,
\end{equation} 
and the statement follows by taking the limit of $\delta$ goes to zero.\qed

\section{Direct quantization and homotopy invariance conditions}

Here we establish that the quantization and homotopy invariance of the non-commutative Chern number both hold for smooth deformations of the lattice-model itself (as opposed to deformations of $p$), as long as the Fermi level is in a region of localized spectrum, characterized by the Aizenman-Molchanov bound on the fractional-powers of the Green's function \cite{Aizenmann1993uf}:
\begin{equation}\label{FracMom1}
\int_\Omega d\mu(\omega) |(h-\epsilon_F)^{-1}(\omega,{\bm x})|^s \leq C_s e^{-s \mathfrak{\beta} |{\bm x}|}.
\end{equation}
Here, $s$ is any positive number strictly smaller than one, $\beta$ is a strictly positive parameter which generally depends on $\epsilon_F$, and $C_s$ is a constant that generally depends on $s$. The symbol $| \cdot |$ on the left hand side, and throughout this section, denotes the matrix norm on ${\cal M}_{Q \times Q}$. 

The technique based on the fractional powers of the Green's function is one of the most effective tools in the analysis of the localization problem. The bound of Eq.~\ref{FracMom1} has been established for all cases where the localization is known to occur \cite{Aizenman1998bf}, such as at large disorder strength \cite{Aizenmann1993uf} or at the edges of the energy spectrum \cite{AizenmanRMP1994hg}. Furthermore, the  bound can be established algorithmically, in a finite number of steps \cite{AizenmanIM2006nv}. This means one can use a computer \cite{ProdanPreparation1} to explore the localization problem beyond the typical situations mentioned above.  We mention that all the characteristics of the localization phenomenon, such as the dynamical localization of the time-evolution operator, spectral localization (i.e the pure point nature of the energy spectrum) or the exponential decay of the eigenstates and of the projector onto the occupied electron states, follow from the bound on the fractional powers of the Green's function \cite{AizenmanIM2006nv}. 

\begin{figure}
\center
  \includegraphics[width=7cm]{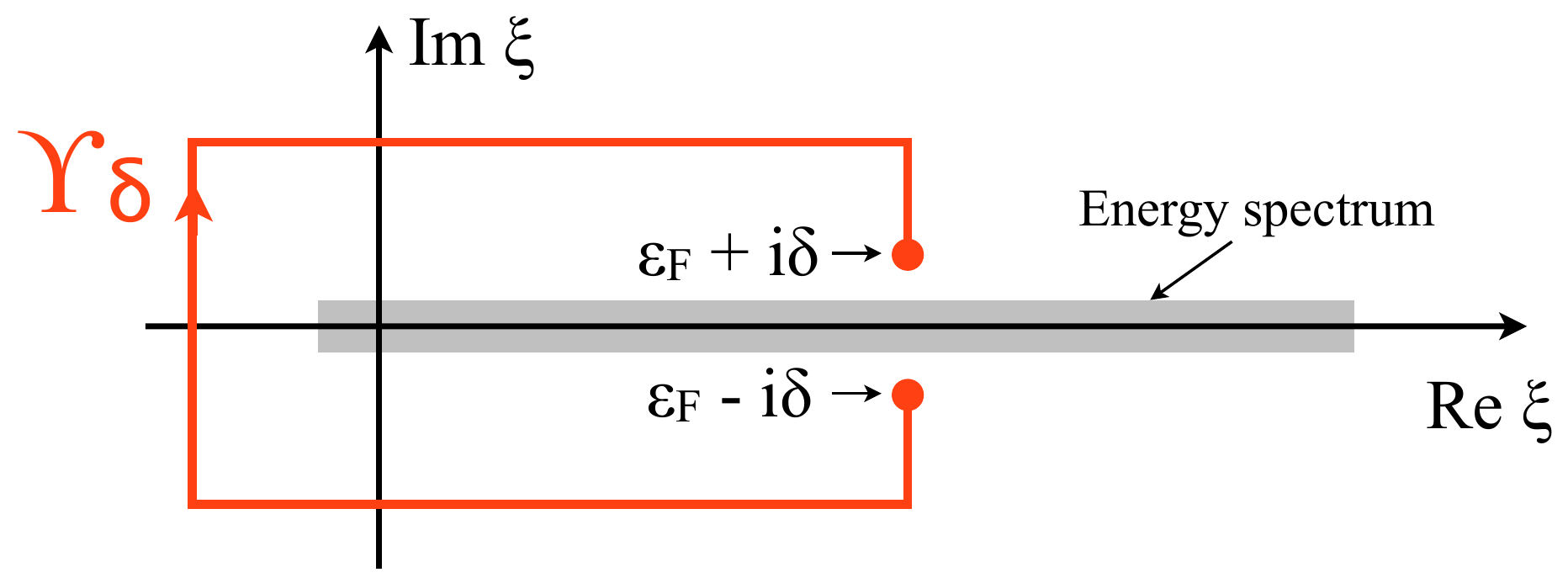}\\
  \caption{The contour $\gamma_\delta$ used in the contour-integral representation of the spectral projector.}
  \label{contour}
\end{figure} 

We will use the bound on the fractional powers of the Green's function in the following way. First, let us note that with the assumption of localization at $\epsilon_F$, the spectral projector $p=\chi_{(-\infty,\epsilon_F]}(h)$ is equal to $\chi_{(-\infty,\epsilon_F)}(h)$ and can be represented as:
\begin{equation}
p=\lim_{\delta \searrow 0} \frac{\imath}{2\pi} \int_{\gamma_\delta} d\xi \ (h-\xi)^{-1},
\end{equation}
where the contour $\gamma_\delta$ is illustrated in Fig.~\ref{contour} and the limit is in the strong topology. Since the limit as $\delta \searrow 0$ of $(h-\epsilon_F \pm \imath \delta)^{-1}(\omega,{\bm x})_{\alpha,\beta}$ ( $= \langle {\bm 0},\alpha |(H_\omega -\epsilon_F \pm \imath \delta)^{-1}|{\bm x},\beta\rangle $) exists for almost all values of $\epsilon_F$ on the real axis (see the technical comments in \cite{AizenmanIM2006nv}), we can formally write the above contour-integral representation as:
\begin{equation}
p=\frac{\imath}{2\pi}\int_{\gamma_{0^+}} d\xi \ (h-\xi)^{-1},
\end{equation}
without any confusion \cite{Aizenman1998bf}. Now, since the resolvent $(h-\xi)^{-1}$ decays exponentially when $\xi$ is away from the real axis, we can always extend the bound \ref{FracMom1} uniformly to the entire $\gamma_{0^+}$:
\begin{equation}\label{FracMom2}
\int_\Omega d\mu(\omega) |(h-\xi)^{-1}(\omega,{\bm x})|^s \leq C_s e^{-s \beta |{\bm x}|}, \ \mbox{for all} \  \xi\in \gamma_{0^+}.
\end{equation}
We now can formulate our statement in precise terms.

\medskip \noindent {\bf Proposition 6.} Let $h$ be the random lattice Hamiltonian defined in Eq.~\ref{LatticeHamOmega}.

\noindent \begin{enumerate}
\item Assume that the bound of Eq.~\ref{FracMom2} holds (hence the energy spectrum near and at the Fermi level is localized). Then $\|p\|_W < \infty$ and the localization length $\Lambda_n$ defined in Eq.~\ref{LocLength} is finite. As such, the non-commutative Chern number takes quantized values.
\item Let $h'=h+\delta h$ be a deformation of the Hamiltonian $h$ induced by a continuous change of the hopping amplitudes of $h_0$, of the Fermi energy and of the disorder strength. Take the contour $\gamma_{0^+}$ large enough so that it surrounds the occupied energy spectrum during the entire deformation, and assume that the bound of Eq.~\ref{FracMom2} holds uniformly during the deformation (hence the spectrum near and at the Fermi level stays localized). Then 
\begin{equation}
\|p'-p\|_W \leq ct. \times \big (\overline{\delta h}\big )^s,
\end{equation}
for any $s < \frac{1}{4n}$, where $\overline{\delta h}=\sup_{\omega,{\bm x}}|\delta h(\omega,{\bm x})|$. In other words, the deformation of the model generates a continuous homotopy between $p$ and $p'$ in the topology induced by the Sobolev norm $\| \ \|_W$. As such, the non-commutative Chern number remains constant and quantized during such deformations.
\end{enumerate}

\medskip\noindent {\it Proof.}  We will use the following simple estimate: If $f_1$, $\ldots$, $f_{2n}$ are elements from the algebra ${\cal A}$, then:
\begin{equation}\label{U1}
|\mathrm{tr}\{(f_{2n}*\ldots *f_1)(\omega,{\bm 0})\}| \leq Q \sum_{{\bm x}_j's \in \mathbb{Z}^{2n}}\prod_{j=1}^{2n}|f_j(\mathfrak{t}_{{\bm x}_j}^{-1}\omega,{\bm x}_{j-1}-{\bm x}_j)|,
\end{equation}
with ${\bm x}_0$ and ${\bm x}_{2n}$ fixed at the origin (which is imposed throughout this section).

\noindent (i) We consider only the terms of the Sobolev norm that contain derivations, because the remaining term can be treated similarly.  We take $f_j = \partial_i p$ for all $j$'s in Eq.~\ref{U1} (with $i$ fixed), and observe that:
\begin{equation}
|f_j (\omega,{\bm x})|\leq \frac{|{\bm x}|}{2 \pi} \int_{\gamma_{0^+}} |d \xi| \ |\mathrm{Im} \ \xi|^{-1+s} \big |(h-\xi)^{-1}(\omega,{\bm x}) \big |^s.
\end{equation}
Then:
\begin{eqnarray}
{\cal T}\left ( \big|\partial_i p\big|^{2n} \right ) 
 \leq \left (\nicefrac{1}{2 \pi} \right )^{2n}Q \int_{\gamma_{0^+}} |d \xi_1| \ |\mathrm{Im} \ \xi_1|^{-1+s} \ldots \int_{\gamma_{0^+}} |d \xi_{2n}| \ |\mathrm{Im} \ \xi_{2n}|^{-1+s} \nonumber \\ 
\indent \times \sum_{{\bm x}_j's \in \mathbb{Z}^{2n}}\int_\Omega d\mu(\omega)   \prod_{j=1}^{2n}|{\bm x}_{j-1}-{\bm x}_{j}|\big |(h-\xi_j)^{-1}(\mathfrak{t}_{{\bm x}_j}^{-1}\omega,{\bm x}_{j-1}-{\bm x}_j)\big|^s. \nonumber
\end{eqnarray}
Holder inequality enables us to continue:
\begin{eqnarray}
\ldots \leq \left (\nicefrac{1}{2 \pi} \right )^{2n}Q \int_{\gamma_{0^+}} |d \xi_1| \ |\mathrm{Im} \ \xi_1|^{-1+s} \ldots \int_{\gamma_{0^+}} |d \xi_{2n}| \ |\mathrm{Im} \ \xi_{2n})|^{-1+s} \\ 
 \times \sum_{{\bm x}_j's \in \mathbb{Z}^{2n}} \prod_{j=1}^{2n}|{\bm x}_{j-1}-{\bm x}_{j}| \left [ \int_\Omega d\mu(\omega)   |(h-\xi_j)^{-1}(\mathfrak{t}_{{\bm x}_j}^{-1}\omega,{\bm x}_{j-1}-{\bm x}_j)|^{2ns}\right ]^{\frac{1}{2n}}. \nonumber
\end{eqnarray}
If we take $s<\frac{1}{2n}$, we can use the bound in Eq~\ref{FracMom2} on the fractional powers of the resolvent. Noting that the remaining integrals over $\xi_j$'s are convergent, the inequality reduces to:
\begin{eqnarray}
{\cal T}\left ( \big|\partial_i p \big |^{2n} \right ) \leq ct. \sum_{{\bm x}_j's \in \mathbb{Z}^{2n}} \prod_{j=1}^{2n} |{\bm x}_{j-1}-{\bm x}_j| e^{-s \beta |{\bm x}_{j-1}-{\bm x}_j|}, \nonumber
\end{eqnarray}
and the remaining sums are evidently convergent.

\noindent (ii) For all $j$'s, we take 
\begin{equation}
f_j=\partial_i(p'-p)=\frac{\imath}{2\pi}\int_{\gamma_{0^+}} d\xi \ \partial_i\big ( (h'-\xi)^{-1}- (h-\xi)^{-1}\big )
\end{equation}
in Eq.~\ref{U1}, and observe again that:
\begin{equation}
\big |f_j(\omega,{\bm x})\big |\leq  \frac{|{\bm x}|}{2\pi} \int_{\gamma_{0^+}} |d \xi| \ |\mathrm{Im} \ \xi |^{-1+s} \big |\big ( (h'-\xi)^{-1}- (h-\xi)^{-1}\big )(\omega,{\bm x})\big |^s.
\end{equation}
Using the resolvent identity, the condition $|\delta h(\omega,{\bm x})| \leq \overline{\delta h} \ \chi_R({\bm x})$, and the generic inequality $|a1+a_2+\ldots |^s \leq |a_1|^s + |a_2|^s \ldots$, we can continue:
\begin{eqnarray}
\ldots \leq  \frac{(\overline{\delta h})^s}{2\pi} |{\bm x}|\sum_{{\bm y},{\bm z}} \chi_R({\bm y}-{\bm z}) \int_{\gamma_{0^+}} |d \xi| \ |\mathrm{Im} \ \xi|^{-1+s} \\
   \indent \indent \times \big |(h'-\xi)^{-1}(\omega,{\bm y})\big |^s  \big |(h-\xi)^{-1}(\mathfrak{t}_{\bm z}^{-1}\omega,{\bm x}-{\bm z})\big |^s.\nonumber
\end{eqnarray}
Then, from Eq.~\ref{U1} we obtain:
\begin{eqnarray}
\indent \indent  {\cal T}\left ( \big|\partial_i (p'-p) \big |^{2n} \right ) \leq \\
 \frac{(\overline{\delta h})^{2ns}}{(2\pi)^{2n}}\int_{\gamma_{0^+}} |d \xi_1| \ |\mathrm{Im} \ \xi_1|^{-1+s} \ldots \int_{\gamma_{0^+}} |d \xi_{2n}| \ |\mathrm{Im} \ \xi_{2n}|^{-1+s} \nonumber \\
 \times \sum_{({\bm x}_j,{\bm y}_j,{\bm z}_j)'s}  \int_\Omega d\mu(\omega)   \prod_{j=1}^{2n}|{\bm x}_{j-1}-{\bm x}_{j}|  \chi_R({\bm y}_j-{\bm z}_j)   \nonumber \\
 \times \big |(h'-\xi_j)^{-1}(\mathfrak{t}_{{\bm x}_j}^{-1}\omega,{\bm y}_j)\big |^s \big |(h-\xi_j)^{-1}(\mathfrak{t}_{{\bm x}_j+{\bm z}_j}^{-1}\omega,{\bm x}_{j-1}-{\bm x}_j-{\bm z}_j)\big |^s. \nonumber
\end{eqnarray}
Holder inequality enables us to continue as:
\begin{eqnarray}
\ldots \leq \frac{(\overline{\delta h})^{2ns}}{(2\pi)^{2n}}\int_{\gamma_{0^+}} |d \xi_1| \ |\mathrm{Im} \ \xi_1|^{-1+s} \ldots \int_{\gamma_{0^+}} |d \xi_{2n}| \ |\mathrm{Im} \ \xi_{2n}|^{-1+s} \nonumber \\
 \indent \times \sum_{({\bm x}_j,{\bm y}_j,{\bm z}_j)'s} \prod_{j=1}^{2n} |{\bm x}_{j-1}-{\bm x}_{j}|  \chi_R({\bm y}_j-{\bm z}_j) \\
 \indent \times \left [ \int_\Omega d\mu(\omega)\big |(h'-\xi_j)^{-1}(\mathfrak{t}_{{\bm x}_j}^{-1}\omega,{\bm y}_j)\big |^{4ns} \right ]^{\frac{1}{4n}}  \nonumber \\
\indent \times \left [ \int_\Omega dP(\omega) \big |(h-\xi_j)^{-1}(\mathfrak{t}_{{\bm x}_j+{\bm z}_j}^{-1}\omega,{\bm x}_{j-1}-{\bm x}_j-{\bm z}_j)\big |^{4ns} \right ]^{\frac{1}{4n}} \nonumber
\end{eqnarray}
If $s<\frac{1}{4n}$, then we can use the bound in Eq~\ref{FracMom2} on the fractional powers of the resolvent, in which case:
\begin{equation}
\ldots \leq ct. \big ( \overline{\delta h} \big )^{2ns} \sum_{({\bm x}_j,{\bm y}_j,{\bm z}_j)'s}  
\prod_{j=1}^{2n}|{\bm x}_{j-1}-{\bm x}_{j}|\chi_R({\bm y}_j-{\bm z}_j) e^{-s \beta(|{\bm y}_j|+|{\bm x}_{j-1}-{\bm x}_j-{\bm z}_j)|)},
\end{equation}
and the remaining sums are evidently convergent. The statement follows. \qed

\ack We acknowledge extremely useful discussions with Hermann Schulz-Baldes. This work was supported by the U.S. NSF grants DMS-1066045, DMR-1056168 and DMS-1160962.

\bibliographystyle{iopart-num}
 
 \medskip \noindent {\bf References:}\smallskip

\bibliography{../../../TopologicalInsulators}

\providecommand{\newblock}{}
\begin{thebibliography}{10}
\expandafter\ifx\csname url\endcsname\relax
  \def\url#1{{\tt #1}}\fi
\expandafter\ifx\csname urlprefix\endcsname\relax\def\urlprefix{URL }\fi
\providecommand{\eprint}[2][]{\url{#2}}

\bibitem{BELLISSARD:1994xj}
Bellissard J, van Elst A and Schulz-Baldes H 1994 {\em J. Math. Phys.\/} {\bf
  35} 5373--5451

\bibitem{ThoulessPRL1982vh}
Thouless D~J, Kohmoto M, Nightingale M~P and den Nijs M 1982 {\em Phys. Rev.
  Lett.\/} {\bf 49} 405--408

\bibitem{Prodan2010ew}
Prodan E, Hughes T and Bernevig B 2010 {\em Phys. Rev. Lett.\/} {\bf 105}
  115501

\bibitem{ProdanJPhysA2011xk}
Prodan E 2011 {\em J. Phys. A: Math. Theor.\/} {\bf 44} 113001

\bibitem{Prodan:2009oh}
Prodan E 2009 {\em Phys. Rev. B\/} {\bf 80} 125327

\bibitem{Prodan2011vy}
Prodan E 2011 {\em Phys. Rev. B\/} {\bf 83} 195119

\bibitem{XuPRB2012vu}
Xu Z, Sheng L, Xing D~Y, Prodan E and Sheng D~N 2012 {\em Phys. Rev. B\/} {\bf
  85} 075115

\bibitem{Moore:2007ew}
Moore J~E and Balents L 2007 {\em Phys. Rev. B\/} {\bf 75} 121306

\bibitem{Fu:2007vs}
Fu L and Kane C~L 2007 {\em Phys. Rev. B\/} {\bf 76} 045302

\bibitem{Hsieh:2008vm}
Hsieh D, Qian D, Wray L, Xia Y, Hor Y~S, Cava R~J and Hasan M~Z 2008 {\em
  Nature\/} {\bf 452} 970

\bibitem{ZHassanRevModPhys2010du}
Hasan M~Z and Kane C~L 2010 {\em Rev. Mod. Phys.\/} {\bf 82} 3045--3067

\bibitem{QiRMP2011tu}
Qi X~L and Zhang S~C 2011 {\em Rev. Mod. Phys.\/} {\bf 83} 1057--1110

\bibitem{QiPRB2008ng}
Qi X~L, Hughes T~L and Zhang S~C 2008 {\em Phys. Rev. B\/} {\bf 78} 195424

\bibitem{Fu:2007ti}
Fu L, Kane C~L and Mele E~J 2007 {\em Phys. Rev. Lett.\/} {\bf 98} 106803

\bibitem{Roy:2009am}
Roy R 2009 {\em Phys. Rev. B\/} {\bf 79} 195322

\bibitem{LeungPRB2012vb}
Leung B and Prodan E 2012 {\em Phys. Rev. B\/} {\bf 85} 205136

\bibitem{LeungJPA2012er}
Leung B and Prodan E 2012 {\em J. Phys. A: Math. and Theor.\/} {\bf 46} 085205

\bibitem{EssinPRL2009bv}
Essin A~M, Moore J~E and Vanderbilt D 2009 {\em Phys. Rev. Lett.\/} {\bf 102}
  146805

\bibitem{Hughes2010gh}
Hughes T, Prodan E and Bernevig B~A 2011 {\em Phys. Rev. B\/} {\bf 83} 245132

\bibitem{EssinPRB2010ls}
Essin A~M, Turner A~M, Moore J~E and Vanderbilt D 2010 {\em Phys. Rev. B\/}
  {\bf 81} 205104

\bibitem{MalashevichNJP2010bv}
Malashevich A, Souza I, Coh S and Vanderbilt D 2010 {\em New J. Phys.\/} {\bf
  12} 053032

\bibitem{Schulz-BaldesCMP2013gh}
Schulz-Baldes H and Teufel S 2013 {\em Commun. Math. Phys.\/} {\bf 319}
  649--681

\bibitem{ProdanAMRX2013bn}
Prodan E 2013 {\em Appl. Math. Res. eXpress\/} {\bf 2013} 176--255

\bibitem{Aizenmann1993uf}
Aizenman M and Molchanov S 1993 {\em Comm. Math. Phys.\/} {\bf 157} 245--278

\bibitem{CONNES:1985cc}
Connes A 1985 {\em Publications Mathematiques de l'I.H.E.S.\/} {\bf 62}
  257--360

\bibitem{Richter2001jg}
Richter T and Schulz-Baldes H 2001 {\em J. Math. Phys.\/} {\bf 42} 3439

\bibitem{Connes:1994wk}
Connes A 1994 {\em Noncommutative Geometry\/} (San Diego, CA: Academic Press)

\bibitem{ConnesGFA1995re}
Connes A and Moscovici H 1995 {\em Geom. Funct. Anal.\/} {\bf 5} 174--243

\bibitem{BellissardLN2003bv}
Bellissard J 2003 {\em Geometric and Topological Methods for Quantum Field
  Theory\/} (River Edge, NJ: World Sci. Publ.) pp 86--156

\bibitem{KrausPRL2012hh}
Kraus Y~E, Lahini Y, Ringel Z, Verbin M and Zilberberg O 2012 {\em Phys. Rev.
  Lett.\/} {\bf 109} 106402

\bibitem{Gomez-LeonPRL2013dd}
Gomez-Leon A and Platero G 2013 {\em Phys. Rev. Lett.\/} {\bf 110} 200403

\bibitem{AvronCMP1989fj}
Avron J~E, Sadun L, Segert J and Simon B 1989 {\em Comm. Math. Phys.\/} {\bf
  124} 595--727

\bibitem{PeierlsZPhys1933fj}
Peierls R~E 1933 {\em Z. f\"ur Phys.\/} {\bf 80} 763--791

\bibitem{BellissardLNP1986jf}
Bellissard J 1986 {\em Lecture Notes in Physics\/} vol 257 ed Dorlas T,
  Hugenholtz M and Winnink M (Springer-Verlag) pp 99--156

\bibitem{SteinAMM1966bv}
Stein P 1966 {\em The American Mathematical Monthly\/} {\bf 73} 299--301

\bibitem{Dixmier1966jc}
Dixmier J 1966 {\em C. R. Acad. Schi. Paris\/} {\bf 262} A1107--A1108

\bibitem{SegalAM1953bv}
Segal I~E 1953 {\em Ann. Math.\/} {\bf 57} 401--457

\bibitem{KosakiJFA1984gf}
Kosaki H 1984 {\em J. Func. Analysis\/} {\bf 59} 123--131

\bibitem{Aizenman1998bf}
Aizenman M and Graf G~M 1998 {\em J. Phys. A: Math. Gen.\/} {\bf 31} 6783--6806

\bibitem{AizenmanRMP1994hg}
Aizenman M 1994 {\em Rev. Math. Phys.\/} {\bf Special Issue} 1163--1182

\bibitem{AizenmanIM2006nv}
Aizenman M, Elgart A, Naboko S, Schenker J~H and Stolz G 2006 {\em Invent.
  Math.\/} {\bf 163} 343--413

\bibitem{ProdanPreparation1}
Prodan E   (in preparation)

\end{thebibliography}

\end{document}